\documentclass[
 reprint,
showpacs,preprintnumbers,
 amsmath,amssymb,
 aps,
prc,
superscriptaddress]{revtex4-1}

\usepackage{graphicx}% Include figure files
\usepackage{dcolumn}% Align table columns on decimal point
\usepackage{bm}% bold math

\begin{document}

\preprint{APS/123-QED}

\title{On the nature of the mass-gap object in the GW190814 event  }% Force line breaks with \\

\author{Luiz L. Lopes}
\email{llopes@cefetmg.br}
 %\altaffiliation[ ]{Universidade Federal de Florianopolis - Santa Catarina - Brazil}%Lines break automatically or can be forced with \\

\affiliation{%
 Centro Federal de Educa\c{c}\~ao Tecnol\'ogica de Minas Gerais Campus VIII; CEP 37.022-560, Varginha - MG - Brasil
}%

\author{Debora P. Menezes}
\affiliation{%
 Departamento de Física, CFM - Universidade Federal de Santa Catarina;  C.P. 476, CEP 88.040-900, Florianópolis, SC, Brasil }

\date{\today}% It is always \today, today,

\begin{abstract}
 In this work, we make a very extensive study on the conditions that
 allow  the mass-gap object in  the GW190814 event to be faced as a
 degenerate star instead of a black hole. We begin revisiting some
 parametrizations of the Quantum Hadrodynamics (QHD) and then study
 under which conditions the hyperons are present in such a massive
 star. Afterwards, using a vector MIT based model, we study if
 self-bound quark stars, satisfying the Bodmer-Witten conjecture
 fulfills all the observational constraints. Finally, we study hybrid
 stars within a Maxwell construction and check for what values of the
 bag, as well as the vector interaction,  a quark core star with only
 nucleons, and with nucleons admixed with hyperons can reach at least 
2.50 M$_\odot$.  We conclude that, depending on the choice of
parameters, none of the possibilities can be completely ruled out, i.e., the
mass-gap object can be a hadronic (either nucleonic or hyperonic), a quark or a hybrid star,
although some cases are more probable than others. 
 \end{abstract}

\maketitle

\section{Introduction}

In 2019, the GW190814 \cite{GW190814} was detected as the result of a
coalescence of a 25.6 $M_\odot$ black-hole and a compact object of
mass 2.5-2.67 $M_\odot$, which can be either the most massive neutron
star or the least massive black hole ever seen, once it lies in the
region known as mass-gap, in between 2.5 and 5 
M$_\odot$.

The possibility of the mass-gap object being a neutron star has
already been studied in previous works. For instance, in ref.~\cite{nuc1}
the influence of the symmetry energy was studied, and a pure nucleonic
neutron star with mass around 2.75 M$_\odot$ was obtained. The
possibility of dark matter admixed in pure nucleonic neutron star was
studied in ref.~\cite{nuc2}, and a maximum mass of 2.50 M$_\odot$ was
shown to be possible. In refs.~\cite{rot1,rot2} the authors discussed the
possibility of hybrid stars and hadronic neutron stars with
hyperons. They have concluded that the mass-gap object can be a hybrid star
only if the star is fast rotating~\cite{rot1}. Nevertheless, the
presence of hyperons seems to be unlikely even in the Kepler frequency
limit~\cite{rot1,rot2}. Using the generic constant-sound-speed
(CSS) parametrization, ref.~\cite{Han} shows that a hybrid star with maximum mass
above 2.50 M$_\odot$ is possible, even in the static case.

On the other hand, ref.~\cite{nopt} rules out
the possibility of quark-hadron phase transition, at least in the
non-rotating case. But, according to a model-independent analysis
based on the sound velocity, quark-cores are indeed expected inside massive
stars \cite{Annala}. $\Delta$-admixed neutron stars were studied in ref.~\cite{Delta}  and can account for the mass-gap object only in the
Keplerian limit, ruling out static stars with $\Delta$ resonances.
In these studies, no static neutron star with exotic matter fulfills the
mass-gap object constraint.

Another possibility is the  mass-gap object being a self-bound strange
star satisfying the Bodmer-Witten conjecture~\cite{Bod,Witten}. This
possibility was studied in  refs.~\cite{ss1,zhang}, where the authors
consider a color superconducting quark matter, and also  in
ref.~\cite{ss2}, where the authors consider a repulsive bag model with 
dynamically generated gluon mass.
Moreover, using CFL NJL model, ref.~\cite{Odilon} also produces quark
stars with masses above 2.50 M$_\odot$.

In this work, we study if the mass gap object in the GW190814 event can be
a purely nucleonic neutron star, a nucleonic admixed hyperon neutron star, 
a self-bound quark star satisfying the Bodmer-Witten
conjecture~\cite{Bod,Witten}, a hybrid star with nucleons and a quark
core or a hybrid star with nucleons, hyperons and a quark core. The paper is organized as follows:

In sec.~\ref{sec2}, we present the Quantum Hadrodynamics (QHD) model
that simulates the interaction between nucleons as well hyperons. We
show how to construct the equation of state (EoS) and discuss some
parametrizations in the light of both, the GW190814 event, as well as
the necessary constraints that satisfy symmetric nuclear matter
properties. We then discuss how hyperons affect the EoS and in what
conditions hyperonic neutron stars can still reach at least 2.50 M$_\odot$.
Constraints related to the radius and tidal deformability of the canonical
1.4 M$_\odot$ neutron stars are also discussed.

In sec.~\ref{sec3}, the vector MIT bag model is presented and we discuss the
possibility of very massive self-bound quark stars. Two
parametrizations for the quark-quark interaction are 
discussed: an universal coupling, where the vector field of all the quarks has the same strength, and a coupling
deduced from symmetry group arguments. The results are again compared
with the constraints imposed by the radius and  the tidal deformability of the canonical star.

In sec.~\ref{sec4} we construct a hybrid branch stability window; i.e,
we obtain values for the bag and for the vector field, where a stable
hybrid star with a mass above 2.50 M$_\odot$ is possible. We also calculate the  size and the  mass of the quark core of these hybrid stars. At the end, the conclusions are drawn.

\section{The QHD model} \label{sec2}

To simulated the interaction between baryons in dense  
cold matter, we use an extended version of the   QHD model, whose
Lagrangian density reads~\cite{Serot,IUFSU}:

\begin{eqnarray}
\mathcal{L}_{QHD} = \bar{\psi}_B[\gamma^\mu(i\partial_\mu  - g_{B\omega}\omega_\mu   - g_{B\rho} \frac{1}{2}\vec{\tau} \cdot \vec{\rho}_\mu)+ \nonumber \\
- (M_B - g_{B\sigma}\sigma)]\psi_B  -U(\sigma) +   \nonumber   \\
  + \frac{1}{2}(\partial_\mu \sigma \partial^\mu \sigma - m_s^2\sigma^2) - \frac{1}{4}\Omega^{\mu \nu}\Omega_{\mu \nu} + \frac{1}{2} m_v^2 \omega_\mu \omega^\mu+  \nonumber \\
 + \frac{1}{2} m_\rho^2 \vec{\rho}_\mu \cdot \vec{\rho}^{ \; \mu} - \frac{1}{4} {P}^{\mu \nu} \cdot {P}_{\mu \nu} + \mathcal{L}_{\omega\rho} + \mathcal{L}_{\phi} , \label{s1} 
\end{eqnarray}
in natural units. $\psi_B$  is the baryonic  Dirac field, where $B$
can stand either for nucleons only ($N$) or can run over  nucleons ($N$) and hyperons ($H$). The $\sigma$, $\omega_\mu$ and $\vec{\rho}_\mu$ are the mesonic fields, while $\vec{\tau}$ are the Pauli matrices.
 The $g's$ are the Yukawa coupling constants that simulate the strong interaction, $M_B$ is the baryon mass,  $m_s$, $m_v$  and $m_\rho$ are
 the masses of the $\sigma$, $\omega$ and $\rho$ mesons respectively.
  The $U(\sigma)$ is the self-interaction term introduced in
  ref.~\cite{Boguta}:

  \begin{eqnarray}
U(\sigma) =  \frac{1}{3}\lambda \sigma^3 + \frac{1}{4}\kappa \sigma^4, \nonumber 
  \end{eqnarray}
 and $\mathcal{L}_{\omega\rho}$ is a non-linear
  $\omega$-$\rho$ coupling interaction as in ref.~\cite{IUFSU}:

\begin{eqnarray}
 \mathcal{L}_{\omega\rho} = \Lambda_{\omega\rho}(g_{N\rho}^2 \vec{\rho^\mu} \cdot \vec{\rho_\mu}) (g_{N\omega}^2 \omega^\mu \omega_\mu) , \label{EL2}
\end{eqnarray}

\noindent which is necessary to correct the slope of the symmetry
energy ($L$) and has a strong influence on the radii and tidal deformation of the neutron stars~\cite{Rafa2011,dex19jpg};
$\mathcal{L}_\phi$ is related  the strangeness hidden $\phi$ vector
meson, which couples only with the hyperons ($H$), not affecting the
properties of symmetric  nuclear matter:

\begin{equation}
\mathcal{L}_\phi = g_{H \phi}\bar{\psi}_H(\gamma^\mu\phi_\mu)\psi_H + \frac{1}{2}m_\phi^2\phi_\mu\phi^\mu - \frac{1}{4}\Phi^{\mu\nu}\Phi_{\mu\nu} , \label{EL3} 
\end{equation}

\noindent as pointed in ref.~\cite{Lopes2020a,Weiss1}, this vector
channel is crucial to obtain massive hyperonic neutron stars. 

As neutron stars are stable macroscopic objects, we need to describe a neutral, chemically
stable matter and hence, leptons are added as free Fermi gases, whose
Lagrangian density is the usual one.

\begin{equation}
 \mathcal{L}_l =  \bar{\psi}_l[\gamma^\mu(i\partial_\mu  - m_l)]\psi_l. \nonumber 
\end{equation}

To solve the equations of motion, we use the mean field approximation (MFA), where the meson fields are replaced by their expectation values. Applying the Euler-Lagrange formalism, and using the quantization rules ($E = \partial^0$ , $k = i\partial^j$) we easily obtain the eigenvalue for the energy:

\begin{equation}
 E_B = \sqrt{k^2 + M_B^{*2}} + g_{B\omega}\omega_0 + g_{B\phi}\phi_0 + \frac{\tau_{3B}}{2}g_{B\rho}\rho_0 , \label{EL4}
\end{equation}
where $M^{*}_B~\equiv~M_B - g_{B\sigma}\sigma_0$ is the effective
baryon mass and $\tau_{3B}$   assumes the value of +1 for p, $\Sigma^+$,  and $\Xi^0$; zero for $\Lambda^0$ and
  $\Sigma^0$; and -1 for n, $\Sigma^{-}$  and $\Xi^-$.
For the leptons, we have:

\begin{equation}
 E_l =  \sqrt{k^2+m_l^2}, \label{EL5}
\end{equation}
and the mesonic fields in MFA are given by:

\begin{eqnarray}
 m_\sigma^2\sigma_0 + \lambda\sigma_0^2 +\kappa\sigma_0^3 = \sum_B g_{B\sigma}n_B^s , \nonumber \\
(m_\omega^2 + 2\Lambda_v\rho_0^2)\omega = \sum_B g_{B\omega}n_B ,
\nonumber \\
m_\phi^2\phi_0  = \sum_B g_{B \phi} n_B, \label{EL6} \\
(m_\rho^2 +2\Lambda_v\omega_0^2)\rho_0 = \sum g_{B\rho}\frac{\tau_{3B}}{2}n_B, \nonumber
\end{eqnarray}
where $\Lambda_v~\equiv~\Lambda_{\omega\rho}g_{N\omega}^2g_{N\rho}^2$, and 
$n_B^s$ and $n_B$ are, respectively, the scalar density and  the
number density of the baryon $B$. Finally, applying Fermi-Dirac
statistics to baryons and leptons and with the help of Eq.~(\ref{EL6}), we can write the total energy density as~\cite{IUFSUtheory}:

\begin{eqnarray}
 \epsilon = \sum_B \frac{1}{\pi^2}\int_0^{k_{Bf}} dk k^2 \sqrt{k^2 + M_B^{*2}} + \nonumber \\
 +\frac{1}{2}m_\sigma^2\sigma_0^2 + \frac{1}{2}m_\omega^2\omega_0^2 + \frac{1}{2}m_\phi^2\phi_0^2 + \frac{1}{2}m_\rho^2\rho_0^2 + \nonumber \\
 +U(\sigma_0) +3 \Lambda_v\omega_0^2\rho_0^2 
 + \label{EL7} \\ + \sum_l \frac{1}{\pi^2}\int_0^{k_{lf}} dk k^2 \sqrt{k^2 + m_l^{2}} , \nonumber
\end{eqnarray}
and the pressure is easily obtained by thermodynamic relations: $p =
\sum_f \mu_f n_f - \epsilon$, where the sum runs over all the
fermions and $\mu_f$ is the corresponding chemical potential. Now, to determine each particle population, we impose that the matter is $\beta$ stable and has total electric net charge equals
to zero.

\subsection{Parametrization}

Our knowledge of  nuclear physics  took a great leap in the last decade. From nuclear masses analysis~\cite{Wang},  passing through nuclear resonances~\cite{Colo,Reinhard}, and heavy ion collisions (HIC)~\cite{Pagano}; we are able to constraint five  parameters of the symmetric nuclear matter at the saturation point: the saturation density itself ($n_0$), the effective nucleon mass ($M_N^*/M_N$), the incompressibility ($K$), the binding energy per
baryon ($B/A$)~\cite{Glen} and the symmetry energy ($S_0$). The experimental  values of these five parameters are taken  from two extensive review articles, ref.~\cite{Dutra2014,Micaela2017}  and are presented in Tab.~\ref{T1}. 
Besides these five parameters, a sixth one is nowadays
a matter of open discussion: the symmetry energy slope, $L$.
For instance, an upper limit for $L$ of 54.6 MeV, 61.9 MeV and 66 MeV
was presented in  ref.~\cite{Paar,Lattimer2013,Steiner1}
respectively. These values are in strong contrast with recent measurements. For instance,
an upper limit of 117.5 MeV was found in ref.~\cite{Estee}, in a study
about  the spectra of pions in intermediate energy collisions, while
PREX2 results not only presented an upper limit as high as 143 MeV,
but also an inferior limit  of only 69 MeV~\cite{PREX2}. Such high
inferior limit makes PREX2 constraint difficult to reconcile with
the results obtained in ref.~\cite{Paar,Lattimer2013,Steiner1}. In
order to maintain consistency with the other constraints, we have
opted to use here 36 MeV $<~L~<$ 86.8 MeV as a constraint, once its was pointed out in ref.~\cite{Micaela2017}.

\begin{widetext}
\begin{center}
\begin{table}[ht]
\begin{center}
\begin{tabular}{|c|c||c|c|c||c|}
\hline 
  & Parameters & &  Constraints  & This model  \\
 \hline
 $g_{N\sigma}$ & 10.0944  &$n_0$ ($fm^{-3}$) & 0.148 - 0.170 & 0.150 \\
 \hline
  $g_{N\omega}$ & 12.8065   & $M^{*}/M$ & 0.56 - 0.75 & 0.594  \\
  \hline
  $g_{N\rho}$ & 14.441   & $K$ (MeV)& 220 - 260                                          &  258  \\
 \hline
$\kappa$ & -10.8093 (fm$^{-1}$) & $S_0$ (MeV) & 30.0 - 35.0 &  30.7  \\
\hline
$\lambda$ &  -30.1486 & $L$ (MeV) & 36 - 86.8 & 42\\
\hline 
$\Lambda_{\omega\rho}$ &  0.045 & $B/A$ (MeV) & 15.8 - 16.5  & 16.31  \\ 
\hline
\end{tabular}
 
\caption{ Parameters of the modified NL3* model~\cite{NL3d}, utilized in this work and their prediction for the symmetric nuclear matter  at the saturation density; the phenomenological constraints are taken from ref.~\cite{Dutra2014,Micaela2017},} 
\label{T1}
\end{center}
\end{table}
\end{center}
\end{widetext}

It is crystal clear that we need a very stiff equation of state (EoS) in order to obtain at least a 2.50 M$_\odot$ star. There are few QHD models that are able to produce such massive neutron star and still be in
agreement with the constraints presented in Tab.~\ref{T1}. We can
highlight the NL3~\cite{NL3}, which although a little outdated, is
still used in very recent studies~\cite{nuc1,nuc2,NL3R}; its updated version, the so-called NL3*~\cite{NL3d}; the NL-RA1~\cite{NLRA1}; and the recent BigApple parametrization~\cite{BA}. But, of all of these
parametrizations, only the BigApple has a symmetry energy 
slope consistent with the constraint 
presented in ref.~\cite{Micaela2017}, 36 MeV $<~L~<$ 86.8 MeV. 
%However, this is a workable problem. 
Nevertheless, with the help of the non-linear 
$\omega-\rho$ coupling given in Eq.~(\ref{EL2}), we can redefine the symmetry energy slope without  affecting any of the others properties of the symmetric nuclear matter. 
A more serious issue is the incompressibility ($K$). Assuming that
$K$ lies between 240 $\pm$ 20, as pointed out in ref.~\cite{Micaela2017}, only the NL3* and the BigApple
fulfill this constraint. As the BigApple produces a maximum star mass of 2.60
M$_\odot$ and the NL3* has a maximum mass of 2.75 M$_\odot$, we use in this work
a modified version of the NL3*, which includes the $\omega-\rho$ coupling.
All the parameters are displayed in Tab~\ref{T1}.

Once the parametrization of the nuclear matter is settled, we need to
focus on the parametrization of the hyperon-meson coupling constants.
There is little information about the hyperon interaction. One of the few
well known quantity is the $\Lambda^0$ potential depth, $U_\Lambda = -28$ MeV.
The hyperon potential depth is defined as~\cite{Glen}:

\begin{equation}
 U_Y = g_{Y\omega}\omega_0 - g_{Y\sigma}\sigma_0. \label{EL8}
\end{equation}

The potential depth for the $\Sigma$ and the $\Xi$ are known with less
precision. In this work we use the standard value $U_\Sigma =$ +30 MeV
and $U_\Xi = - 4$ MeV, which was recently favored by lattice QCD
calculations~\cite{LQCD}. Nevertheless, the knowledge of the hyperon potential
depth only solves partially the problem once from Eq.~(\ref{EL8}), different
combinations of $g_{Y\omega}$ and $g_{Y\sigma}$ can produce the same
value for the potential depth. And, as pointed in ref.~\cite{Glen},
different values of the hyperon-meson coupling constants can lead to a
maximum mass difference up to 100$\%$. The situation becomes even worse,
as the potential depth give us no information about the hyperon-meson
coupling constants  with the $\rho$ and the $\phi$ mesons.

An alternative is the use of symmetry group arguments. For instance, all coupling constants for the vector mesons can be fixed simultaneous by applying
SU(6) symmetry group~\cite{Lopes2020a,Weiss1}. 
However, in general, SU(6) symmetry group produces neutron stars not
as massive as desired in the present investigation. We can solve this
problem by breaking the SU(6) symmetry group into a more general SU(3)
flavor symmetry group. In this case, the coupling constant for the vector meson depends
of a single parameter, $\alpha$. Therefore we have for the $\omega$ meson ~\cite{IUFSUtheory,lopesnpa,lopesPRC}:

\begin{equation}
\frac{g_{\Lambda\omega}}{g_{N\omega}} = \frac{4 + 2\alpha}{5 + 4\alpha}, \quad \frac{g_{\Sigma\omega}}{g_{N\omega}} = \frac{8 - 2\alpha}{5 + 4\alpha}, \quad 
\frac{g_{\Xi\omega}}{g_{N\omega}} = \frac{5 - 2\alpha}{5 + 4\alpha} ,
\nonumber
\end{equation}

for the $\phi$ meson:

\begin{eqnarray}
 \frac{g_{\Lambda\phi}}{g_{N\omega}} = \sqrt{2} \bigg ( \frac{2\alpha - 5}{5 + 4\alpha} \bigg ) , \quad \frac{g_{\Sigma\phi}}{g_{N\omega}} = -\sqrt{2} \bigg (
 \frac{2\alpha_v + 1 }{5 + 4\alpha} \bigg ) \nonumber ,\\ \frac{g_{\Xi\phi}}{g_{N\omega}} =  - \sqrt{2} \bigg ( \frac{2\alpha + 4}{5 + 4\alpha} \bigg ) , \quad 
 \frac{g_{N\phi}}{g_{N\omega}} = 0 \nonumber,
\end{eqnarray}

and finally for the $\rho$ meson:

\begin{equation}
 \frac{g_{\Sigma\rho}}{g_{N\rho}} = 2\alpha, \quad \frac{g_{\Xi\rho}}{g_{N\rho}} = 
 2\alpha_v - 1 , \quad \frac{g_{\Lambda\rho}}{g_{N\rho}} = 0. \label{EL9}
\end{equation}

\begin{figure*}[t]
\begin{tabular}{cc}
\centering % \begin{center}/\end{center} takes some additional vertical space
\includegraphics[scale=.51, angle=270]{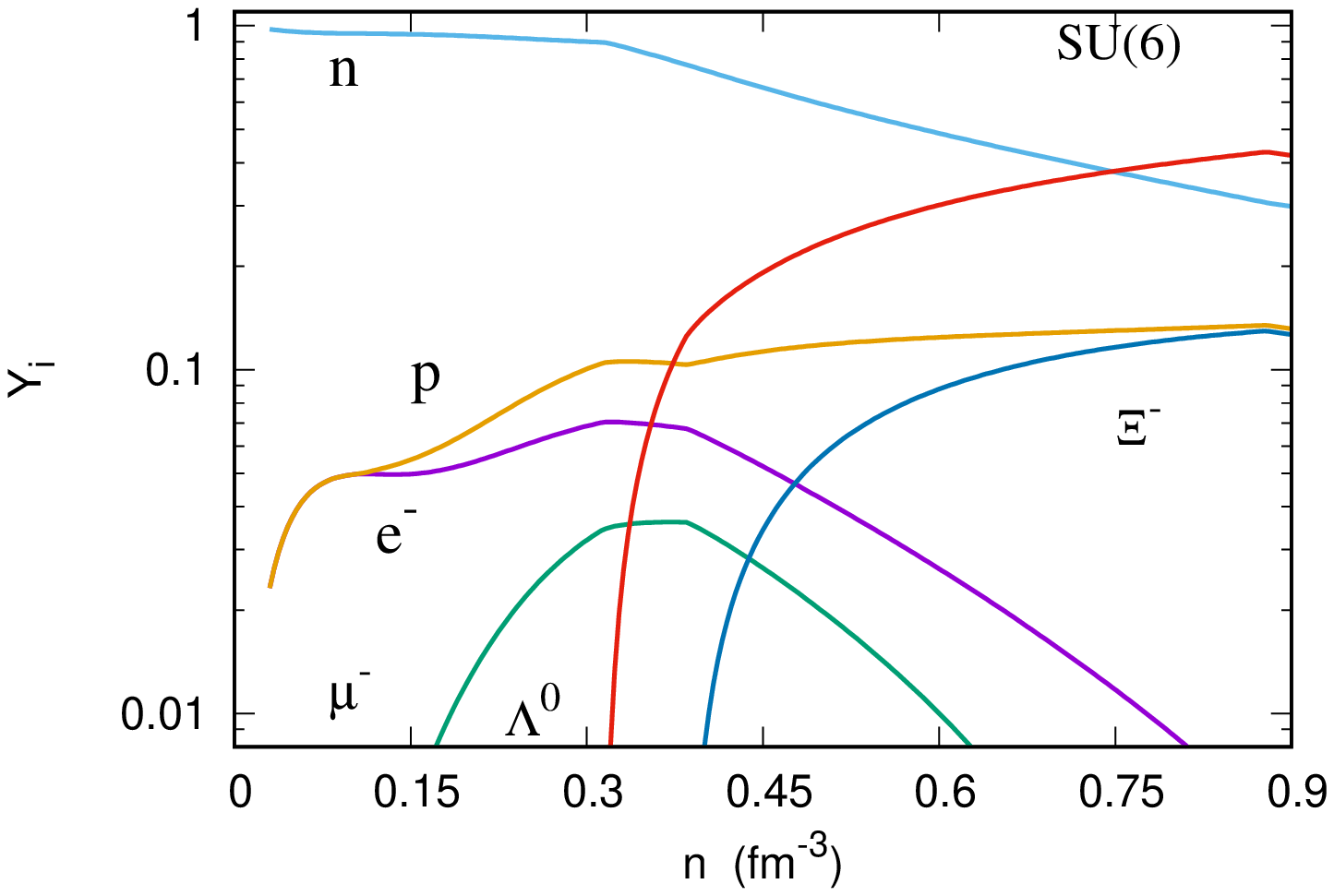} &
\includegraphics[scale=.51, angle=270]{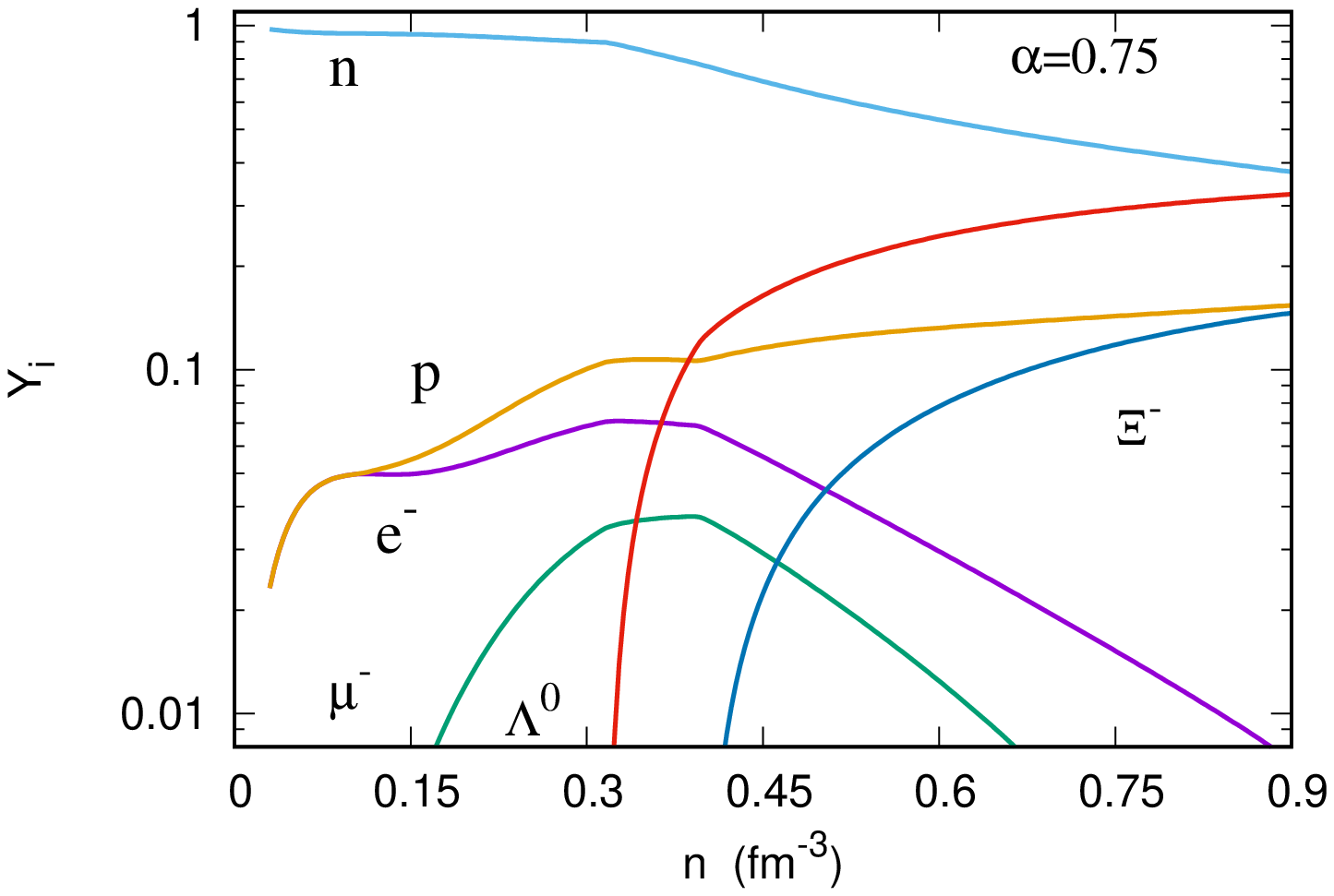}\\
\includegraphics[scale=.51, angle=270]{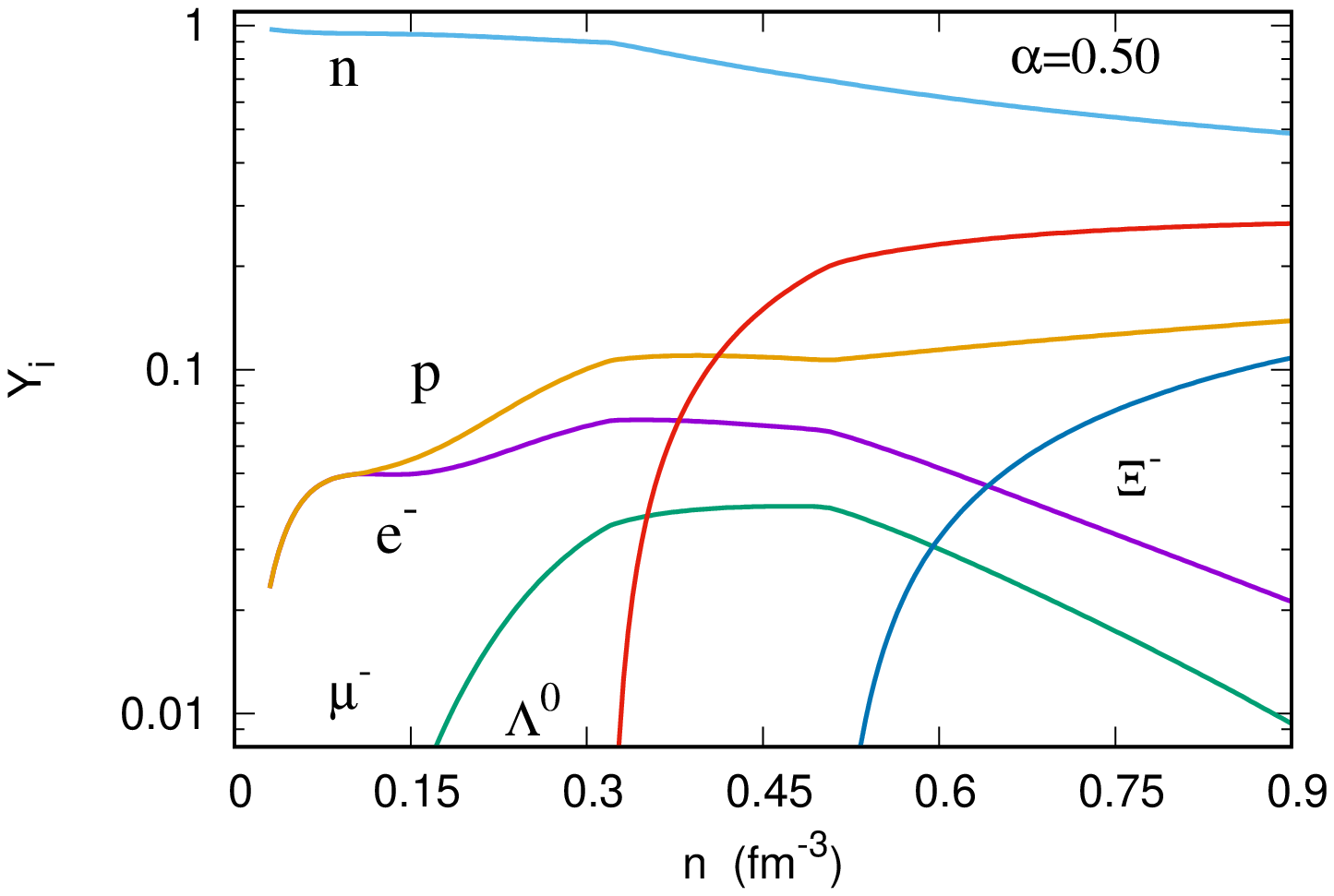} &
\includegraphics[scale=.51, angle=270]{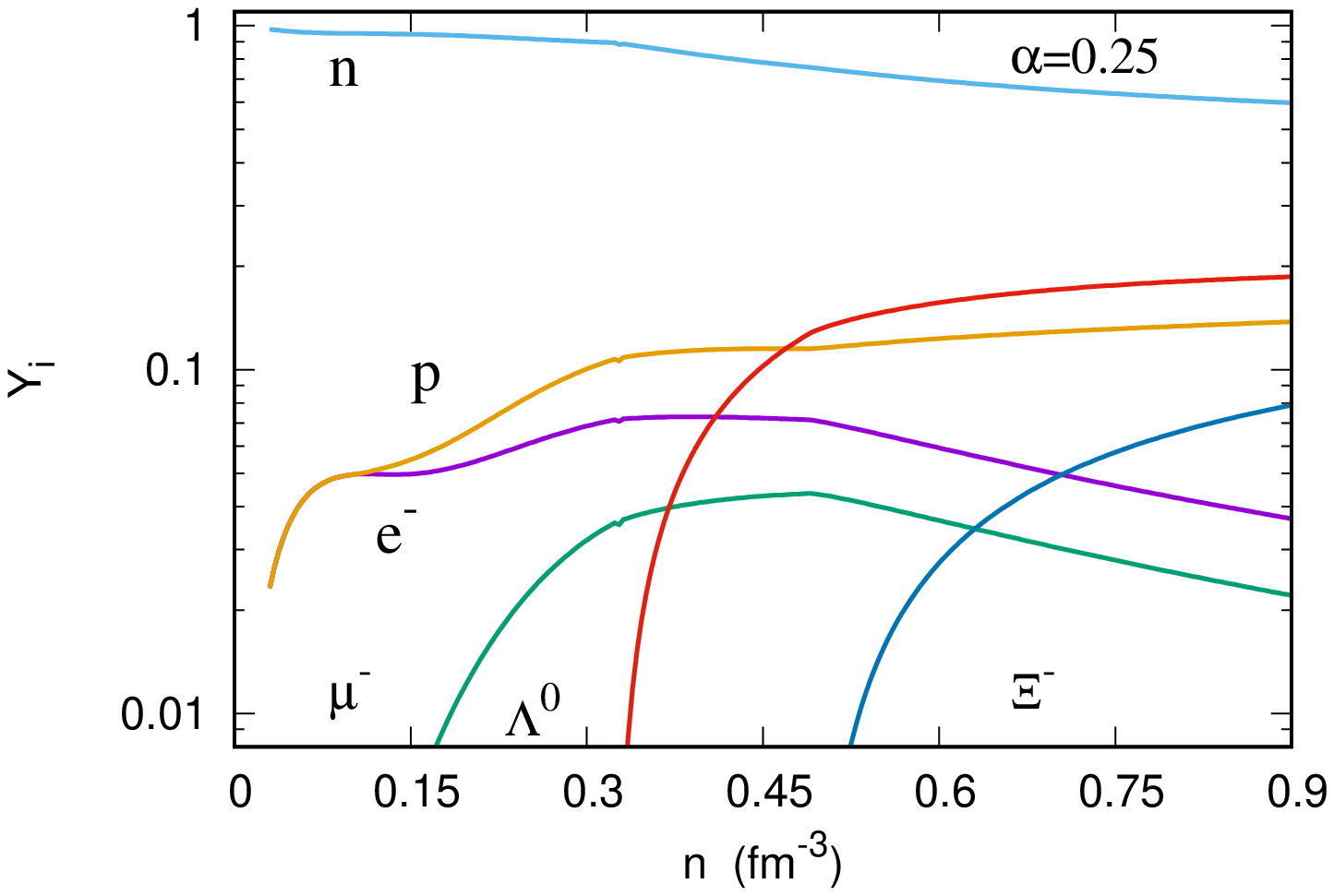}\\
\end{tabular}
% "\includegraphics" is very powerful; the graphicx package is already loaded
\caption{(Color online) Particle population for the SU(6) group and
  different values of $\alpha$. } \label{FL1}
\end{figure*}

We next choose some arbitrary values for $\alpha$ and calculate the
vector meson coupling constants, while the scalar meson coupling
constants are fixed in order to reproduce realistic values of the
potential depth, as mentioned above: $U_\Lambda$  = $-28$ MeV,
$U_\Sigma=$  = $+30$~MeV and $U_\Xi$  = $-4$~MeV.  The calculated
coupling constants  are displayed in Tab.~\ref{TL2}, while  we plot in
Fig.~\ref{FL1} the particle population for these values of $\alpha$.
Notice that for $\alpha$ = 1 the  group SU(6) is recovered.

\begin{table}[ht]
\begin{center}
\begin{tabular}{|c|c|c|c|c|}
\hline
 - & $\alpha = 1.00$ & $\alpha = 0.75$ &  $\alpha = 0.50$ &  $\alpha = 0.25$   \\
\hline
 $g_{\Lambda\omega}/g_{N\omega}$        & 0.667 & 0.687   & 0.714 & 0.75   \\
\hline
 $g_{\Sigma\omega}/g_{N\omega}$         & 0.667 & 0.812  & 1.00 & 1.25   \\
 \hline
$g_{\Xi\omega}/g_{N\omega}$           & 0.333 & 0.437  & 0.571 & 0.75   \\
\hline
$g_{\Lambda\phi}/g_{N\omega}$           & -0.471 & -0.619  & -0.808 & -1.06   \\
\hline
$g_{\Sigma\phi}/g_{N\omega}$           & -0.471 & -0.441  & -0.404 & -0.354   \\
\hline
$g_{\Xi\phi}/g_{N\omega}$           & -0.943 & -0.972  & -1.01 & -1.06   \\
\hline
$g_{\Sigma\rho}/g_{N\rho}$           & 2.0 & 1.5  & 1.0 & 0.5   \\
\hline
$g_{\Xi\rho}/g_{N\rho}$           & 1.0 & 0.5  & 0.0 & -0.5   \\
\hline
$g_{\Lambda\sigma}/g_{N\sigma}$           & 0.613 & 0.629  & 0.651 & 0.679   \\
\hline
$g_{\Sigma\sigma}/g_{N\sigma}$           & 0.461 & 0.578  & 0.730 & 0.930   \\
\hline
$g_{\Xi\sigma}/g_{N\sigma}$           & 0.279 & 0.374 & 0.428 & 0.616   \\
\hline
\end{tabular}
 \caption{Hyperon-meson coupling constants for different values of $\alpha$. When we impose $\alpha$ = 1 we recover the  group SU(6).}\label{TL2}
 \end{center}
 \end{table}
 
 \begin{figure}[htb] 
\begin{centering}
 \includegraphics[angle=270,width=0.47\textwidth]{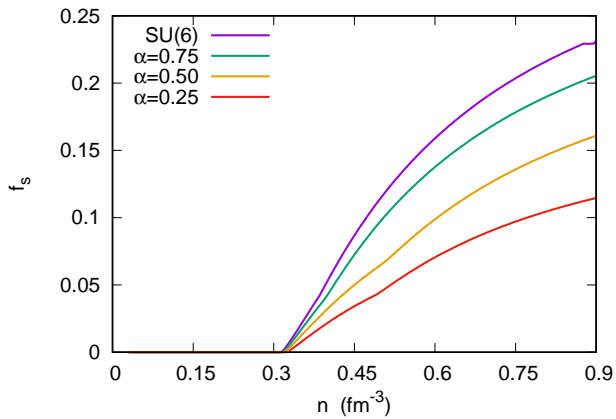}
\caption{(Color online) Strangeness fraction as a function of $\alpha$.} \label{FL2}
\end{centering}
\end{figure}

 We can see that for all values of $\alpha$, the only hyperons present
 are the $\Lambda^0$ and the $\Xi^-$. 
Also, for all values of $\alpha$,  the $\Lambda^0$ is the first
hyperon to appear, at densities 
around 0.33 fm$^{-3}$. For the SU(6) parametrization, the potentials
depth are less repulsive at high density. Due to this fact, the
hyperons are favored, resulting in two interesting features:
the $\Lambda^0$ becomes the most populous particle at densities around
0.73 fm$^{-3}$, matter is practically deleptonized, and the electric charge neutrality is reached with equal proportions of protons and $\Xi^{-}$ hyperons. For $\alpha$ =0.75, the matter is also deleptonized, but at higher densities. For lower values, the potentials depth are already too repulsive, and electrons and muons are always present.

A more clever way to understand the variation of strangeness content particle is, instead of looking at the individual hyperon population, looking at the strangeness fraction, $f_s$, defined as:

\begin{equation}
 f_s = \frac{1}{3} \frac{\sum n_i |s_i|}{n} , \label{EL10}
\end{equation}
where $s_i$ is the strangeness of the $i-th$ baryon. The results are plotted in Fig.~\ref{FL2}.
 As we can see, there is a direct relation between $\alpha$ and the
 strangeness fraction, as also pointed out in ref.~\cite{lopesPRC}. When we move away from SU(6) we increase the repulsion of the hyperons, by increasing the $Y-\omega$ and the $Y-\phi$ coupling constants as shown in Tab.~\ref{TL2}. This reduces the hyperon population at high densities, reducing the strangeness fraction.

Now we use the EoS for different values of $\alpha$ as input to solve the TOV equations~\cite{TOV}:

\begin{eqnarray}
 \frac{dp}{dr} = \frac{-GM(r)\epsilon (r)}{r^{2}} \bigg [ 1 + \frac{p(r}{\epsilon(r)} \bigg ]  \nonumber \\ 
 \times \bigg  [ 1 + \frac{4\pi p(r)r^3}{M(r)} \bigg ] \bigg [ 1 - \frac{2GM(r)}{r} \bigg ]^{-1} , \nonumber \\
 \frac{dM}{dr} =  4\pi r^2 \epsilon(r) . \label{EL11}
\end{eqnarray}

\begin{figure}[ht]
  \begin{centering}
\begin{tabular}{c}
\includegraphics[width=0.333\textwidth,angle=270]{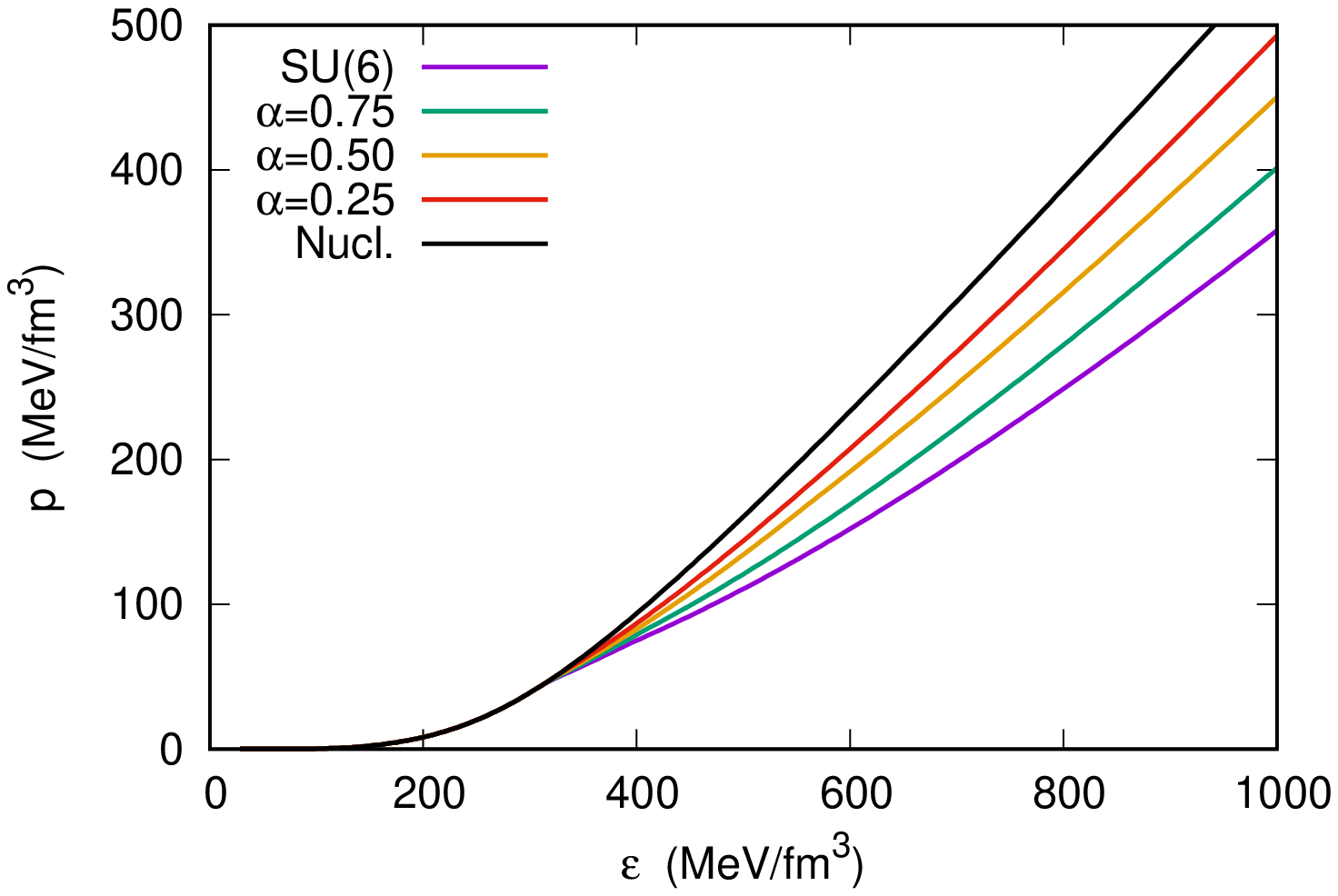} \\
\includegraphics[width=0.333\textwidth,,angle=270]{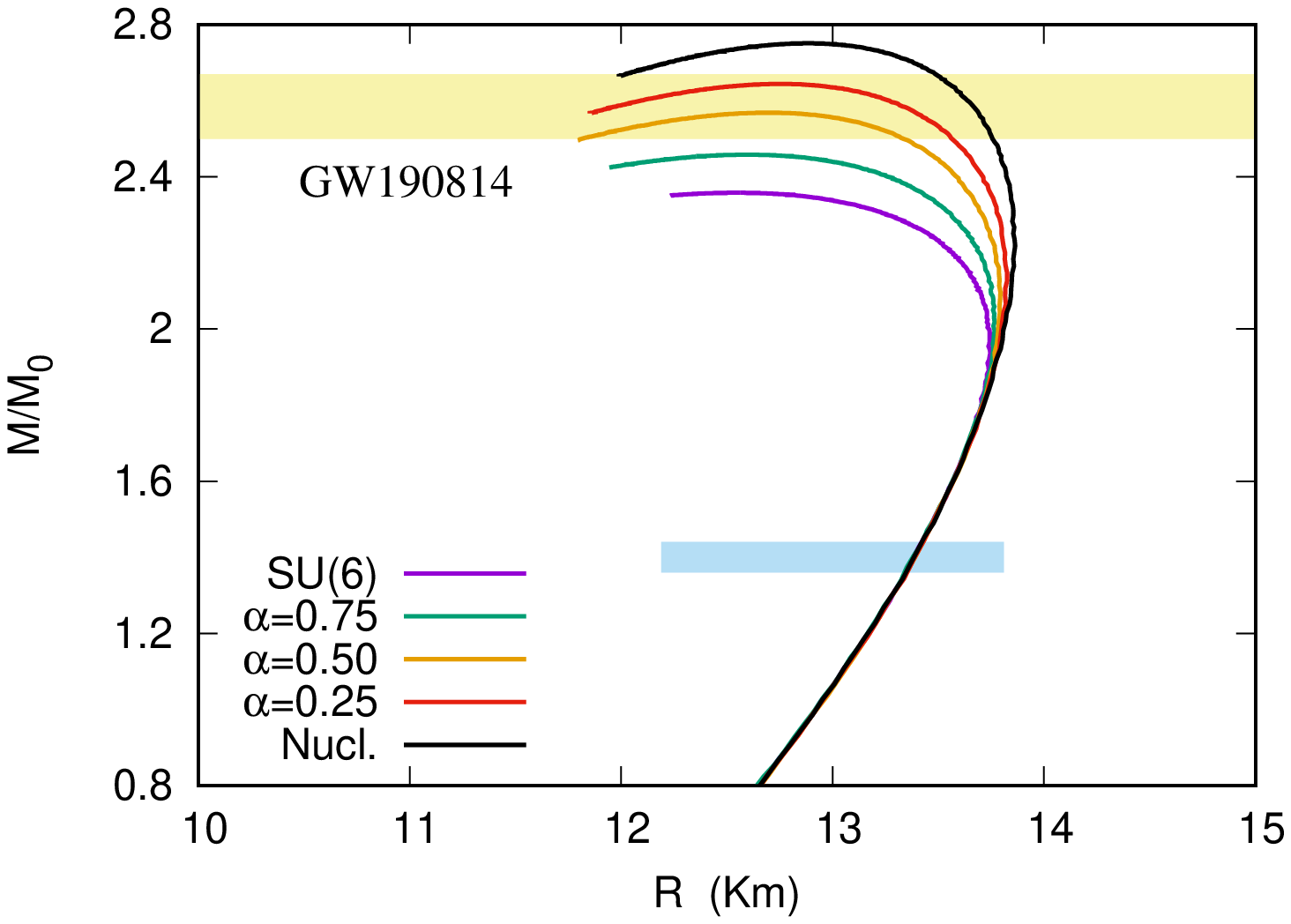} \\
\end{tabular}
\caption{(Color online) Top: EoS for the different values of the parameter $\alpha$.  Bottom: Mass radius relation for pure nucleonic and hyperonic stars. The yellowish hatched area refers   to the uncertainty of the lower mass compact object present in the   GW190814 event, while the blueish hatched area is the uncertainty about the radius of the 1.4M$_\odot$ star. Both constraints are taken from ref.~\cite{GW190814}. The radius of all canonical stars is 13.38 km.} \label{FL3}
\end{centering}
\end{figure}

The EoS as well the mass-radius relation of the TOV solution are displayed 
in Fig.~\ref{FL3}, where we use the BPS EoS to simulate the neutron star crust~\cite{BPS}. As can be seen, there is a clear relation between
the value of $\alpha$, the strangeness fraction, $f_s$, the EoS and
the maximum mass. The lower the value of the $\alpha$, the lower the
value of $f_s$, which produces stiffer EoS, as well as more massive stars. 
In the light of the GW190814 event, we can see that this mass-gap object not only can be a pure
nucleonic neutron star, but also a hyperon admixed star, if $\alpha~<0.75$.
For $\alpha~\geq$ 0.75, the maximum mass lies below 2.50 M$_\odot$. 
Also, as the $\Lambda^0$ onset happens around 0.33 fm$^{-3}$ for all
values of $\alpha$, no hyperonic neutron stars with mass below 1.66
M$_\odot$ is possible. 

We can also discuss our results in the light of some observational 
astronomical constraints. One of the hot topics in modern days is
the radius of the canonical star, M = 1.40 M$_\odot$. However,
as pointed in ref.~\cite{Rafa2011}, the radius of the canonical
star  strongly depends of the symmetry energy slope, $L$. And,
as discussed earlier, the high uncertainty about its value 
will obviously affect the uncertainty of the radius. For instance,
in ref.~\cite{Ozel2016}, a very small upper limit, of only 11.1 km
was presented. Another small value of  11.9 km was also pointed in
ref.~\cite{Capano}, while in ref.~\cite{Steiner2} an upper limit 
of 12.45 km was deduced. More recently,  results obtained from
Bayesian analysis indicate that the radius of the canonical star lies between 
10.8 km and 13.2 km in ref.~\cite{Yuxi}; and 11.3 km to 13.5 km in
ref.~\cite{Michael}; whilst results coming from 
the NICER x-ray telescope points out that  $R_{1.4}$ lies between
11.52 km and 13.85 km from ref.~\cite{NICER1} and between 11.96 km and
14.26 km from ref.~\cite{NICER2}. State-of-the-art theoretical results
at low and high baryon density point to an upper limit of $R_{1.4}$
$<$ 13.6 km~\cite{Annala2}. Finally, PREX2 results~\cite{PREX2} 
indicate that the radius of the canonical star lies between 13.25 km
$<~R_{1.4}~<$ 14.26 km. This constraint  is mutually exclusive with
the result presented in refs.~\cite{Ozel2016,Capano,Steiner2,Yuxi}.

Notwithstanding, in this work we use as a constraint the radius of
the canonical star between 12.2 km $<~R_{1.4}~<$13.7 km, as presented 
in ref.~\cite{GW190814}. The reason why we choose this value over all
the others discussed above is because it is derived from the GW190814 event itself, the subject
of the present work. As, in our study no hyperon is present in a 1.4
M$_\odot$ star, its radius for all values of $\alpha$  is 13.38
km. Such value is in agreement with our main constraint from
ref.~\cite{GW190814}, as well as it is in agreement with NICER
results~\cite{NICER1,NICER2}, Bayesian analysis~\cite{Michael},
PREX2~\cite{PREX2}, and state-of-the-art theoretical
results~\cite{Annala2}. 
However, it is in disagreement with the results presented in ref.~\cite{Ozel2016,Capano,Steiner2,Yuxi}. 

Another important quantity and constraint is the so-called
dimensionless tidal deformability  parameter $\Lambda$. If we put an
extended body in an inhomogeneous external field it will experience
different forces throughout its extent. The result is a tidal
interaction. The tidal deformability of a compact object is a single
parameter $\lambda$ that quantifies how easily the object is deformed
when subjected to an external tidal field. A larger tidal deformability
indicates that the object is easily deformable. On the opposite side,
a compact object with a small tidal deformability parameter is more
compact and more difficult to deform.  The tidal deformability
is defined as the ratio between the induced 
quadrupole $Q_{ij}$ and the perturbing tidal field $\mathcal{E}_{ij}$ that causes the perturbation:

\begin{equation}
 \lambda = - \frac{Q_{ij}}{\mathcal{E}_{ij}} . \label{EL12}
\end{equation}

However, in the literature is more commonly found the dimensionless
tidal deformability parameter $\Lambda$ defined as:

\begin{equation}
 \Lambda~\equiv~\frac{\lambda}{M^5}~\equiv~\frac{2k_2}{3C^5} , \label{EL13}
\end{equation}
where $M$ is the compact object mass and $C = GM/R$ is its
compactness. The parameter $k_2$ is called the second (order) Love number:

\begin{eqnarray}
 k_2 =  \frac{8C^5}{5}(1-2C)^2[2 + 2C(y_R -1) -y_R] \nonumber \\
 \times \{ 2C[6 -3y_R +3C(5y_R -8)] \nonumber \\
 +4C^3[13 -11y_R +C(3y_R -2) +2C^2(1+y_R)] \nonumber \\
 +3(1-2C)^2[2-y_R +2C(y_R -1))]\ln(1-2C)\} ^{-1}, \nonumber
 \\ \label{EL14}
\end{eqnarray}
where $y_R=y(r=R)$ and $y(r)$ is obtained  by solving:

\begin{equation}
 y\frac{dy}{dr} +y^2 + yF(r) +r^2Q(r) = 0 . \label{EL15}
\end{equation}
 Eq.~(\ref{EL15}) must be solved coupled with the TOV equations,
 Eq.~(\ref{EL11}). The coefficients $F(r)$ and $Q(r)$ are given by:

\begin{equation}
 F(r) = \frac{1 -4\pi G r^2[\epsilon(r) - p(r)]}{E(r)} ,
\end{equation}

\begin{eqnarray}
 Q(r) = \frac{4\pi G}{E(r)} \bigg [ 5\epsilon(r) + 9p(r) + \frac{\epsilon(r)+p(r)}{\partial p/ \partial \epsilon} - \frac{6}{4\pi Gr^2} \bigg ]  \nonumber \\
 -4 \bigg [ \frac{G[M(r) + 4\pi r^3 p(r)]}{r^2 E(r)} \bigg ]^2, \nonumber  \\ \label{EL17}
\end{eqnarray}
where $E(r) = (1 -2GM(r)/r)$. Additional discussion about the theory
of the tidal deformability, as well  as the tidal Love numbers are beyond
the scope of this work and can be found in ref.~\cite{Odilon, tidal1,tidal2,tidal3,tidal4} and references therein.

\begin{figure}[ht] 
\begin{centering}
 \includegraphics[angle=270,width=0.45\textwidth]{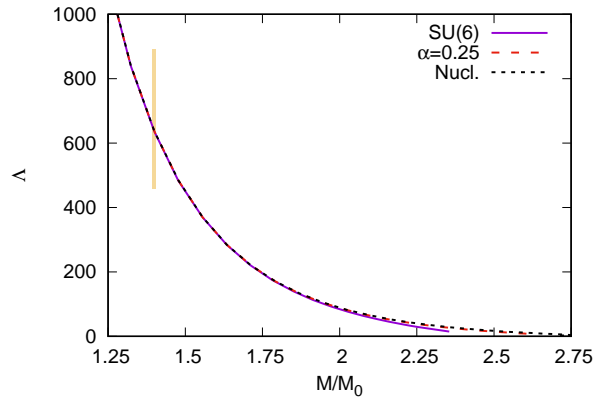}
\caption{(Color online) Dimensionless tidal parameter for pure nucleonic stars, as well for two values of $\alpha$. The hatched area refers   to the uncertainty of the tidal deformability parameter deduced from the    GW190814 event ~\cite{GW190814}.} \label{FL4}
\end{centering}
\end{figure}

We display in Fig.~\ref{FL4} the tidal deformability parameter
$\Lambda$ as a function of the  star mass of pure nucleonic stars as
well as for hyperonic stars with two values of $\alpha$. 
Exactly as in the case of the radius, the dimensionless tidal parameter of the canonical star, $\Lambda_{1.4}$ can be used as a
constraint.  An upper limit  of 860 was found in
ref.~\cite{Michael}. A  close limit, $\Lambda_{1.4}~<$ 800 was pointed in ref.~\cite{tidal4}. In ref.~\cite{Yuxi}, an upper
limit of 686 was deduced from  Bayesian analysis. On the other hand,
two mutually exclusive constraints are presented in
ref.~\cite{tidal1}, which proposed a limit between 70
$<~\Lambda_{1.4}~<$ 580, and  the PREX2 inferred values, whose 
limit lies between 642 $<~\Lambda_{1.4}~<$ 955~\cite{PREX2}. 
Here, as done in the case of the radius, we use the values 458
$<~\Lambda_{1.4}~<$ 889 as a constraint, once it was derived from the
GW190814 event itself and is presented in ref.~\cite{GW190814}. As no
hyperons are present in a 1.4M$_\odot$ star, with the choice of
parameters used in this study, all canonical stars have the same value
for the $\Lambda_{1.4}$ = 644. This value is in agreement with the
main constraint from ref.~\cite{GW190814}, 
as well as with all the others, except for the one presented in ref.~\cite{tidal1}.

Another important  quantity of dense cold nuclear matter is the speed
of sound, defined as:

\begin{equation}
 v_s^2 = \frac{\partial p}{\partial \epsilon} . \label{EL18}
\end{equation}

The speed of sound is related to the stiffness of the EoS, and can give us
important insight about the internal composition of  the neutron star.
In general,  for a pure nucleonic neutron star, the speed of sound
grows monotonically. However, the onset of new degrees of freedom can
produce a non-trivial behaviour and the speed of the sound can present
maxima and minima. The speed of the sound also acts as a constraint, once super-luminal values, $v_s^2 > 1$, violate Lorentz symmetry. Moreover, due to the conformal symmetry of the QCD, perturbative QCD (pQCD) predicts an upper limit $v_s^2 < 1/3$ at very high densities, $n > 40~ n_0$~\cite{Paulo}. Such high density is far beyond those found in the neutron star interiors.

\begin{figure}[ht] 
\begin{centering}
 \includegraphics[angle=270,width=0.45\textwidth]{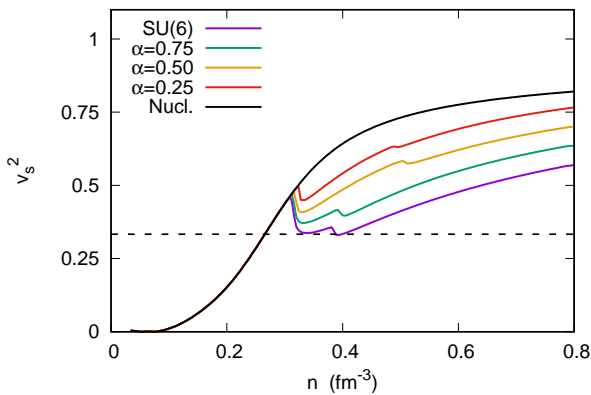}
\caption{(Color online) Square of the speed of sound for pure nucleonic and hyperonic matter for different values of of $\alpha$. The horizontal line is the conformal limit predicted by pQCD.} \label{FL5}
\end{centering}
\end{figure}

The speed of the sound may also be linked to the size of the quark
core of a hybrid star. As recently pointed out in ref.~\cite{Annala},
the speed of the sound of the quark matter is closely related to the mass and radius
of the quark core in hybrid stars. The authors found that if the
conformal bound ($v_s^2 < 1/3$ ) is not strongly violated, massive
neutron stars are predicted to have sizable quark-matter cores.  We
show in the fig.~\ref{FL5} the square of the speed of the sound for
pure nucleonic matter, as well as for hyperonic matter for different values of $\alpha$.

As can be seem, all of our models are causal, $v_s^2 < 1$. This was
expected, since we are dealing with a relativistic model. The onset of
hyperons breaks the monotonic behaviour and reduces the speed of the sound. 
The higher the value of $\alpha$, the lower is the speed of the
sound. Nevertheless, the conformal limit is always violated. 

The last constraint for hadronic neutron stars presented in this work
is the minimum mass that enables the direct Urca (DU) process. As
pointed in ref.~\cite{DU}, any acceptable EoS shall not allow the
direct Urca process to occur in neutron stars with masses below 1.5
M$_\odot$. Therefore, we also use this value as constraint. The trigger to the nucleonic direct Urca channel is directly related to the leptonic fraction $x_{DU}$, defined as~\cite{Rafa2011,DU}:

\begin{equation}
 x_{DU} = \frac{1}{1 +(1 +x_e^{1/3})^3} , \label{EL19}
\end{equation}
where $x_e = n_e/(n_e +n_\mu)$. For the models presented in this work,
only pure nucleonic neutron stars enable DU process, and at very high
density: 0.51 fm$^{-3}$, which imply a very massive star of 2.56
M$_\odot$ as an inferior limit for the DU process.

We now display in Tab.~\ref{TL3} some macroscopic and microscopic
properties of nucleonic and hyperonic neutron stars. As we have
already pointed out, all star families have 
the same radius and tidal deformability for the canonical mass. Also,
we need $\alpha~ <$ 0.75 in order to reproduce a 2.50 M$_\odot$
star. From the microscopic point of view, we can infer some features
from the speed of the sound and the strangeness fraction. For
instance, our study indicates that the central speed of the sound
should be $v_s^2~>0.62$ in order to produce at least a 2.50 M$_\odot$. Such high value is almost twice the conformal limit imposed by pQCD~\cite{Paulo}. In the same sense, we also need
$f_s~<$ 18$\%$ at the core of the neutron star. When we look at the
more well established  2.0 M$_\odot$, we see that a strangeness
fraction of only 3.6$\%$ is enough to prevent a formation of a 2.50 M$_\odot$.

\begin{widetext}
\begin{center}
\begin{table}[ht]
\begin{center}
\begin{tabular}{|c|c|c|c|c|c|c|c|c|c|c}
\hline 
 Model & $M_{max}/M_0$ & R (km) & $n_c$ (fm$^{-3}$) & $f_{sc}$  &  $v_{sc}^2$ & $M_{DU}/M_\odot$ & $R_{1.4}$ (km) & $\Lambda_{1.4}$ & $f_{s2.0}$    \\
 \hline
 SU(6) & 2.36  & 12.56 & 0.777 & 0.210 & 0.56  & - & 13.38 & 644 & 0.056 \\
 \hline
$\alpha$ = 0.75  & 2.46   & 12.59  & 0.762 & 0.181  & 0.62 & - & 13.38 & 644 &  0.036 \\
  \hline
  $\alpha$ = 0.50 & 2.57   & 12.70 & 0.736 & 0.133 & 0.68 & - & 13.38 & 644 & 0.025                            \\
 \hline
$\alpha$ = 0.25 & 2.64  & 12.75 & 0.720 & 0.093  & 0.74 & - & 13.38 & 644 & 0.014  \\
\hline
Nucl. &  2.75 & 12.87 & 0.699 &  - & 0.79 & 2.56 & 13.38 & 644 & -  \\
\hline 
Constraints & $>~2.50$ & - & - & - & $ <~ 1.0$ & $>~1.50$ & $12.2-13.7$  & $458-889$ & -  \\ 
\hline
\end{tabular}
 
\caption{ Nucleonic and hyperonic neutron star properties for different values of $\alpha$ and some astrophysical constraints.} 
\label{TL3}
\end{center}
\end{table}
\end{center}
\end{widetext}

\section{Vector MIT bag model} \label{sec3}

Let's now consider the possibility of the mass-gap object of the GW190814 being a stable quark star, sometimes called strange star.
This idea is based on the so-called Bodmer-Witten
hypothesis~\cite{Bod,Witten}. According to it, the matter composed of
protons and neutrons may be only meta-stable. The true ground state of
strongly interacting matter would therefore consist of strange matter
(SM), which in turn is composed of deconfined up, down and strange
quarks. For the SM hypothesis to be true, the energy per baryon of of the deconfined phase (for p = 0 and T = 0) is lower than the nonstrange infinite baryonic matter. Or explicitly~\cite{Bod,Witten,lopes2021ps1}:

\begin{equation}
 E_{(uds)}/A < 930 ~ \mbox{MeV}, \label{EL20} 
\end{equation}
at the same time, the nonstrange matter still need to have  an energy per baryon higher than the one of nonstrange infinite baryonic matter, otherwise, protons and neutrons would decay into $u$ and $d$ quarks:
\begin{equation}
 E_{(ud)}/A > 930 ~\mbox{MeV} \label{EL21} .
\end{equation}

Therefore, both, eq.~(\ref{EL20}) and (\ref{EL21}) must  simultaneous
be true.

One of the simplest model to study quark matter is to so called MIT
bag model~\cite{MIT}. This model considers that each baryon is
composed of three non-interacting quarks inside a bag. The bag, in
turn, corresponds to an infinity potential that confines the
quarks. In this simple model the quarks are free inside the bag and
are forbidden to reach its exterior. All the information about the
strong force relies on the bag pressure value, which mimics the vacuum
pressure. However, the maximum mass of a stable quark star in the
original MIT bag model is way below 2.50 M$_\odot$.
We can overcome this issue by using a modified MIT bag model which adds a massive repulsive vector field, analogous to the $\omega$ meson of the QHD. We follow this path here. The Lagrangian of the modified MIT bag model reads~\cite{lopes2021ps1}:

\begin{eqnarray}
\mathcal{L}_{vMIT} = \{ \bar{\psi}_q[\gamma^\mu(i\partial_\mu  - g_{qV}V_\mu)  - m_q ]\psi_q + \nonumber \\
+ \frac{1}{2}m_V^2V_\mu V^\mu - B \}\Theta(\bar{\psi}_q\psi_q) . \label{EL22}
\end{eqnarray}
where $m_q$ is the mass of the quark $q$ of flavor $u$, $d$ or $s$.
Here, we follow ref.~\cite{lopes2021ps1} and use ($m_u=m_d=4$ MeV,
$m_s=95$ MeV); $\psi_q$ is the Dirac quark 
field, $B$ is the constant vacuum pressure, and $\Theta(\bar{\psi}_q\psi_q)$ is the Heaviside step function to assure that the quarks exist only confined to the bag.
The quark interaction is mediated by the massive vector channel
$V_\mu$ analogous to the $\omega$ meson in QHD~\cite{Serot}.  Besides,
leptons are added to account for $\beta$ stable matter. Imposing MFA,
and applying Euler-Lagrange to Eq.~(\ref{EL22}), we obtain the energy
eigenvalue for the quark, as well as the expected value for the vector field.

\begin{equation}
 E = \sqrt{m_q^2 + k^2} + g_{qV}V_0, \label{EL23}
\end{equation}
\begin{equation}
 m_V^2V_0 = \sum_q g_{qV} n_q , \nonumber
\end{equation}
 where $n_q$ is the number density of the quark $q$. Now, applying
 Fermi-Dirac statistics, the energy density is analogous to the QHD plus the bag term:

 \begin{eqnarray}
 \epsilon = \sum_q \frac{N_c}{\pi^2}\int_0^{k_f} dk k^2 \sqrt{k^2 + m_q^{2}}  +\frac{1}{2}m_V^2V_0^2  + B \nonumber \\
 + \sum_l \frac{1}{\pi^2}\int_0^{k_f} dk k^2 \sqrt{k^2 + m_l^{2}} , \label{EL24}
\end{eqnarray}
where $N_c$ is the number of colours and $p =\mu n - \epsilon$. 

The interaction of the vector field with different quark flavors can follow two different prescriptions. 
In the first one we have an universal coupling, i.e. the strength of the interaction is the same for all three quarks. This is a more conventional approach, and was done, for instance, in ref.~\cite{Gomes1,Gomes2}. In this case,
$g_{uV} = g_{dV} = g_{sV}$. Another possibility explored in ref.~\cite{lopes2021ps1} is that the vector field is not only analogous to the $\omega$ meson, but it is the $\omega$ meson itself. In this approach we can use symmetry group arguments and 
construct an invariant Lagrangian. In this approach we have: $g_{uV}=g_{dV}$ and $g_{sV}=0.4g_{uV}$. We now follow ref.~\cite{lopes2021ps1} and  define two quantities:

\begin{equation}
 G_V~\equiv~ \bigg (\frac{g_{uV}}{m_V} \bigg )^2 \quad \mbox{and} \quad 
 X_V~\equiv~  \frac{g_{sV}}{g_{uV}} , \label{EL25}
\end{equation}
$G_V$ is related to the absolute strength of the vector field itself; $X_V$ is related to the relative strength of the vector field. If $X_V$ = 1.0 we are dealing with an universal coupling,
while $X_V$ =0.4 implies symmetry group arguments. We study how different values of $G_V$ and $X_V$ affect the macroscopic properties of strange stars.

\begin{figure}[ht] 
\begin{centering}
 \includegraphics[angle=270,width=0.45\textwidth]{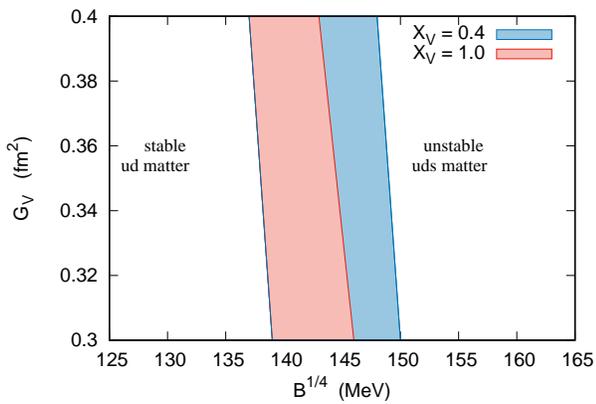}
\caption{(Color online) Stability window for $X_V$ = 1.0 and $X_V = 0.4$} \label{FL6}
\end{centering}
\end{figure}

\begin{figure*}[t]
\begin{tabular}{cc}
\centering % \begin{center}/\end{center} takes some additional vertical space
\includegraphics[scale=.51, angle=270]{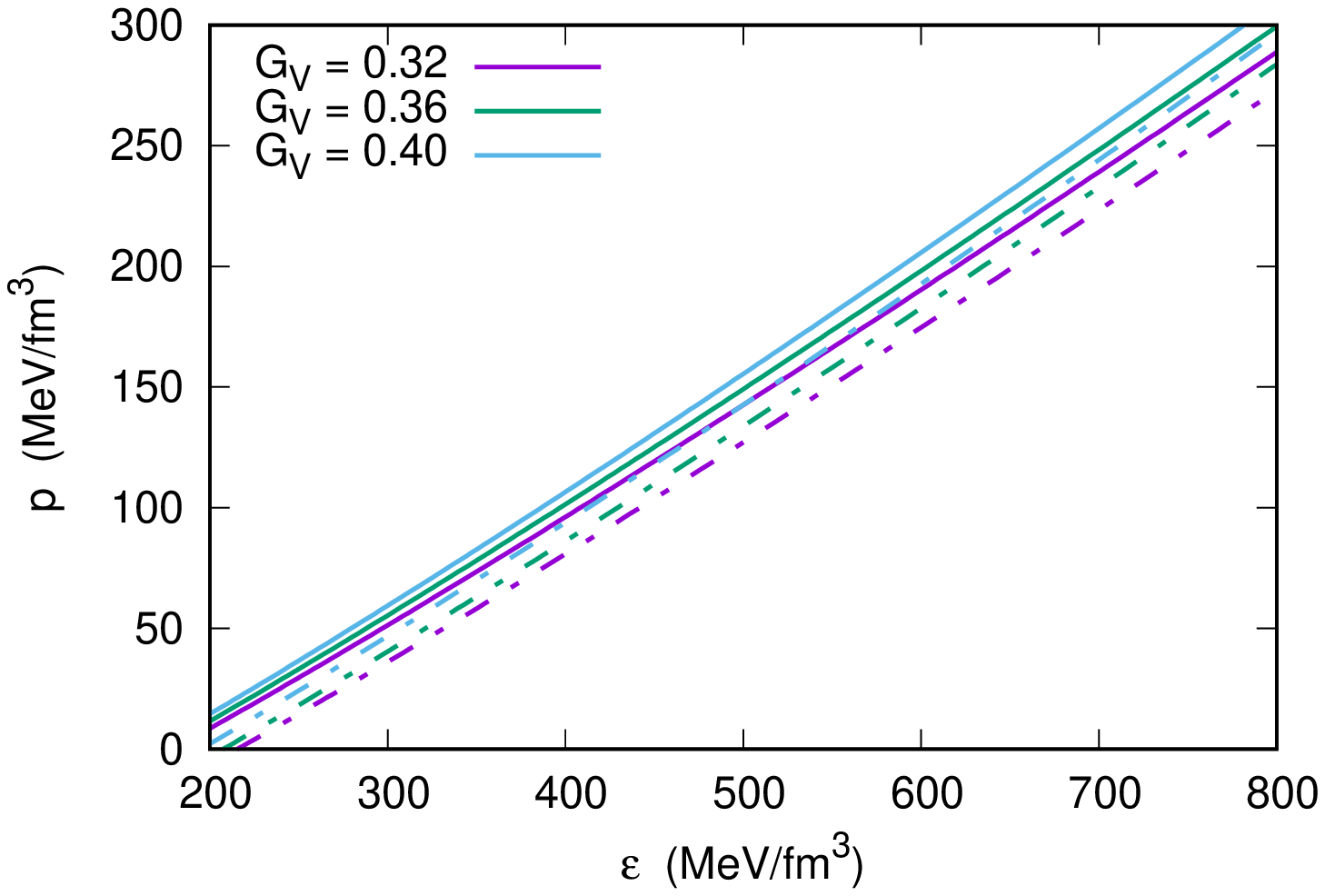} &
\includegraphics[scale=.51, angle=270]{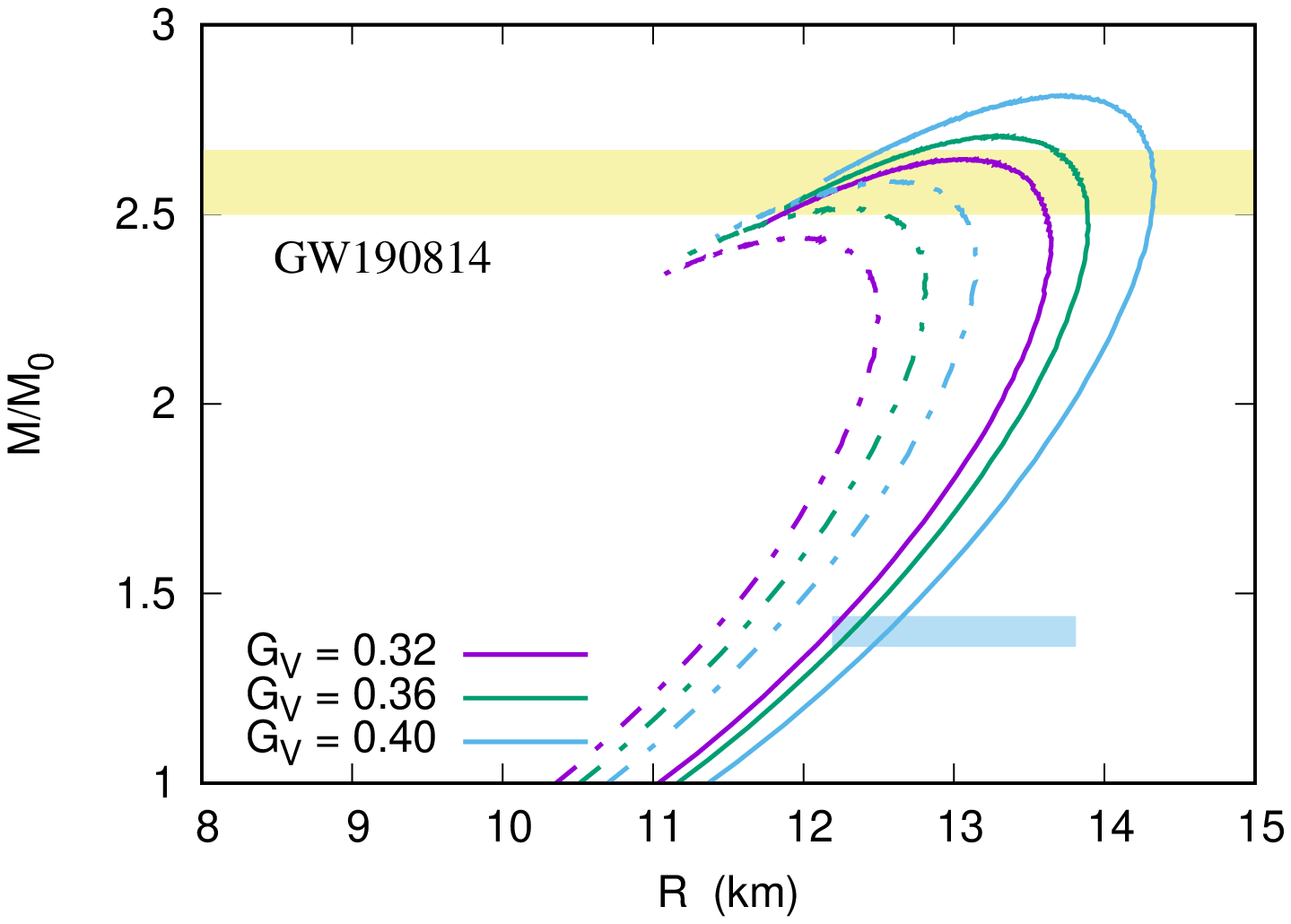}\\
\includegraphics[scale=.51, angle=270]{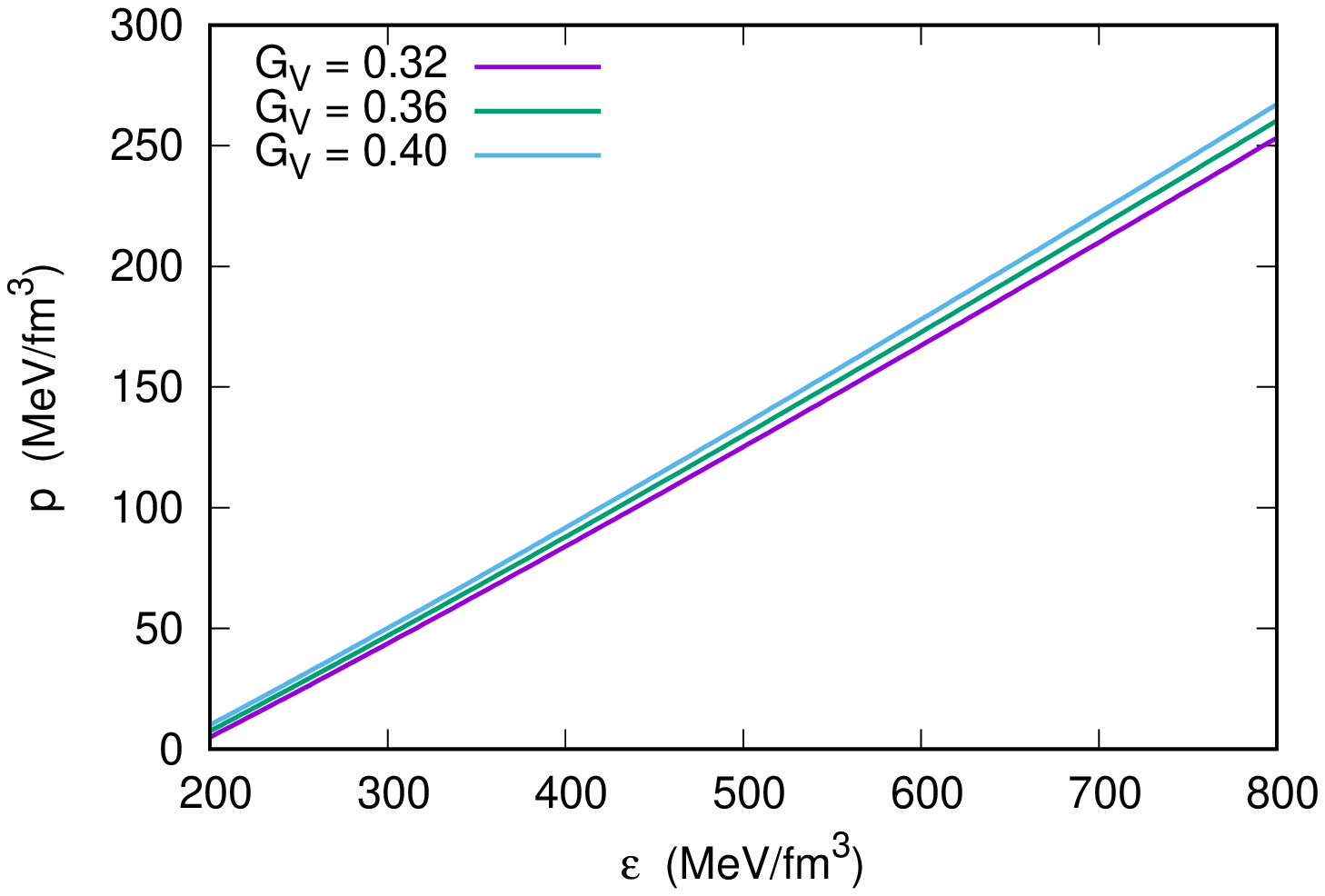} &
\includegraphics[scale=.51, angle=270]{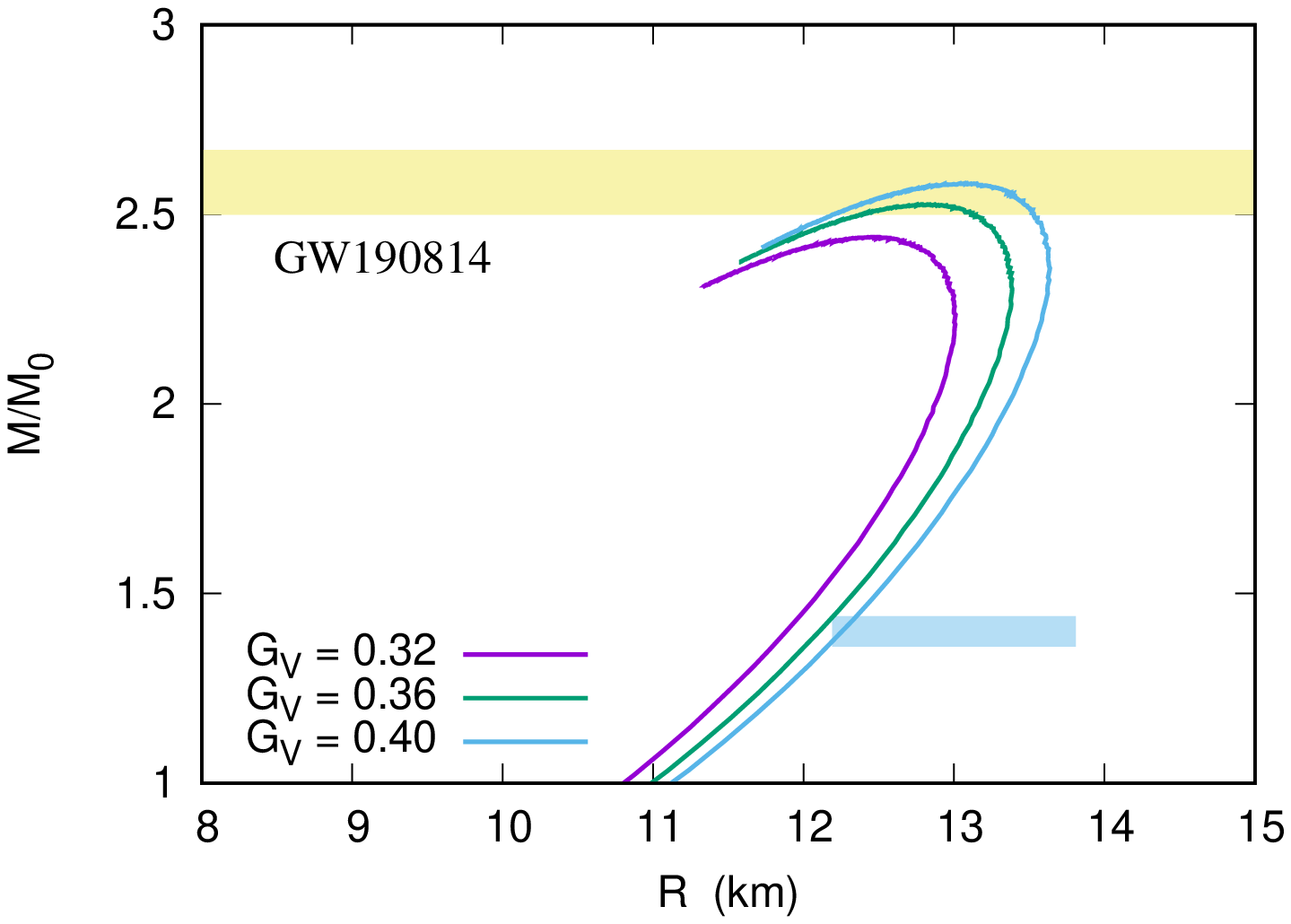}\\
\end{tabular}
% "\includegraphics" is very powerful; the graphicx package is already loaded
\caption{(Color online) EoS (left) and TOV solution (right) for strange stars with $X_V$ = 1.0 (top) and $X_V$ =0.4 (bottom).
Solid lines indicate minimum bag value for the stability window and  dotted lines indicate maximum bag value for the stability window.} \label{FL7}
\end{figure*}

\begin{table}[htb]
\begin{center}
\begin{tabular}{|c|c|c|c|c|}
\hline
$G_V$ (fm$^2$)  &  $X_V$ & $B^{1/4}_{Min}$ (MeV) & $B^{1/4}_{Max}$ (MeV)     \\
\hline
  0.30  & 0.4  & 139  &  150     \\
 \hline
  0.32  & 0.4  & 138 &  149    \\
 \hline
 0.36  & 0.4  & 138 &  148    \\
 \hline
0.40  & 0.4  & 137 &  148    \\
 \hline
 \hline
 0.30 &  1.0  & 139 &  146    \\
 \hline
 0.32  & 1.0  & 138 &  145   \\
 \hline
 0.36  & 1.0  & 138 &  144    \\
 \hline
0.40  & 1.0  & 137 &  143    \\
 \hline
  \end{tabular}
 \caption{Stability windows obtained with the vector
MIT bag model. }\label{TL4}
\end{center}
 \end{table}

Now, for a chosen value of $G_V$ and $X_V$, the bag pressure $B$ is not arbitrary. To predict the existence of strange stars, $B$ must be chosen in order to satisfy both Eq.~(\ref{EL20}) and
Eq.~(\ref{EL21}). The set of values 
that satisfies both equations simultaneous is used to construct the stability window~\cite{Bod,Witten}. 
We display the stability window for 
0.30 fm$^2~<G_V~<~$0.40 fm$^2$ in Fig.~\ref{FL6} and the numerical results are shown in Tab.~\ref{TL4}.

We can see that the lower limit of the stability window is independent of the value of $X_V$. This is expected, once 
it is related to the stability of the two flavored quark matter,
expressed in eq.~(\ref{EL21}). On the other side, the stability of three flavored quark matter depends on the $X_V$. 
The stability window is always wider for $X_V$ =0.4, once the repulsion in the 
strange quark is lower, reducing the energy per baryon.
We now show in Fig.~\ref{FL7} the EoS and the mass-radius relation for strange stars with $X_V$ =1.0 and $X_V$ = 0.4. 
For $X_V=1.0$, we plot the EoS and the TOV solution for both, minimum and maximum values of the bag as presented in Tab.~\ref{TL4}. As can be seen, for a given bag value, the higher the value of $G_V$,
the stiffer is the EoS and consequently, the higher is the maximum mass. For $X_V$ = 1.0, a maximum mass above 2.50 M$_\odot$ is
reached for the minimum bag value for all three values of  $G_V$
presented in this work. In the case of the maximum bag value of
Tab.~\ref{TL4}, we see that only $G_V~>0.32$ can reach at least 2.50
M$_\odot$.  For $X_V$ = 0.4, we plot the results only for the minimum
bag value. We see that the EoS is softer when compared with $X_V$ =
1.0, and therefore produces lower values for the maximum mass, and
only values $G_V~>0.32$ fm$^2$ can reach at least 2.50 M$_\odot$.

Another important constraint is the radius of the 1.4 M$_\odot$
star. Considering only the results that reach at least 
2.50 M$_\odot$, the radii shown in Fig.~\ref{FL7} lie between 11.59 km
and 12.58 km.  We see that several radii agree with the NICER results
~\cite{NICER1,NICER2}, Bayesian analysis~\cite{Michael,Yuxi}, as well
as with ref.~\cite{Annala}. However, as we use here the results coming
from ref.~\cite{GW190814} as the main constraint, not all radius
values are in the range between 12.2 km and 13.7 km. Indeed, neither stars
produced with the maximum bag value nor stars with $G_V$ = 0.32 fm$^2$ have a radius above 12.2 km.

Now we check the values of the dimensionless tidal parameter
$\Lambda$.  As pointed in ref.~\cite{Odilon,tidal5},
the value of $y_R$ in  Eq.~(\ref{EL14}) must be corrected,
since strange stars are self-bound and present a discontinuity
at the surface. Therefore we must have:

\begin{equation}
 y_R \rightarrow y_R - \frac{4\pi R^3 \Delta\epsilon_S}{M} , \label{EL26}
\end{equation}
where $R$ and $M$ are the star radius and mass respectively, and $\Delta\epsilon_S$ is the difference between the energy density at the surface (p =0) and the exterior of the star (which implies $\epsilon=0$). The results for $X_V$=1.0 and $X_V$ = 0.4 are plotted in Fig.~\ref{FL8}:

\begin{figure}[ht]
  \begin{centering}
\begin{tabular}{c}
\includegraphics[width=0.333\textwidth,angle=270]{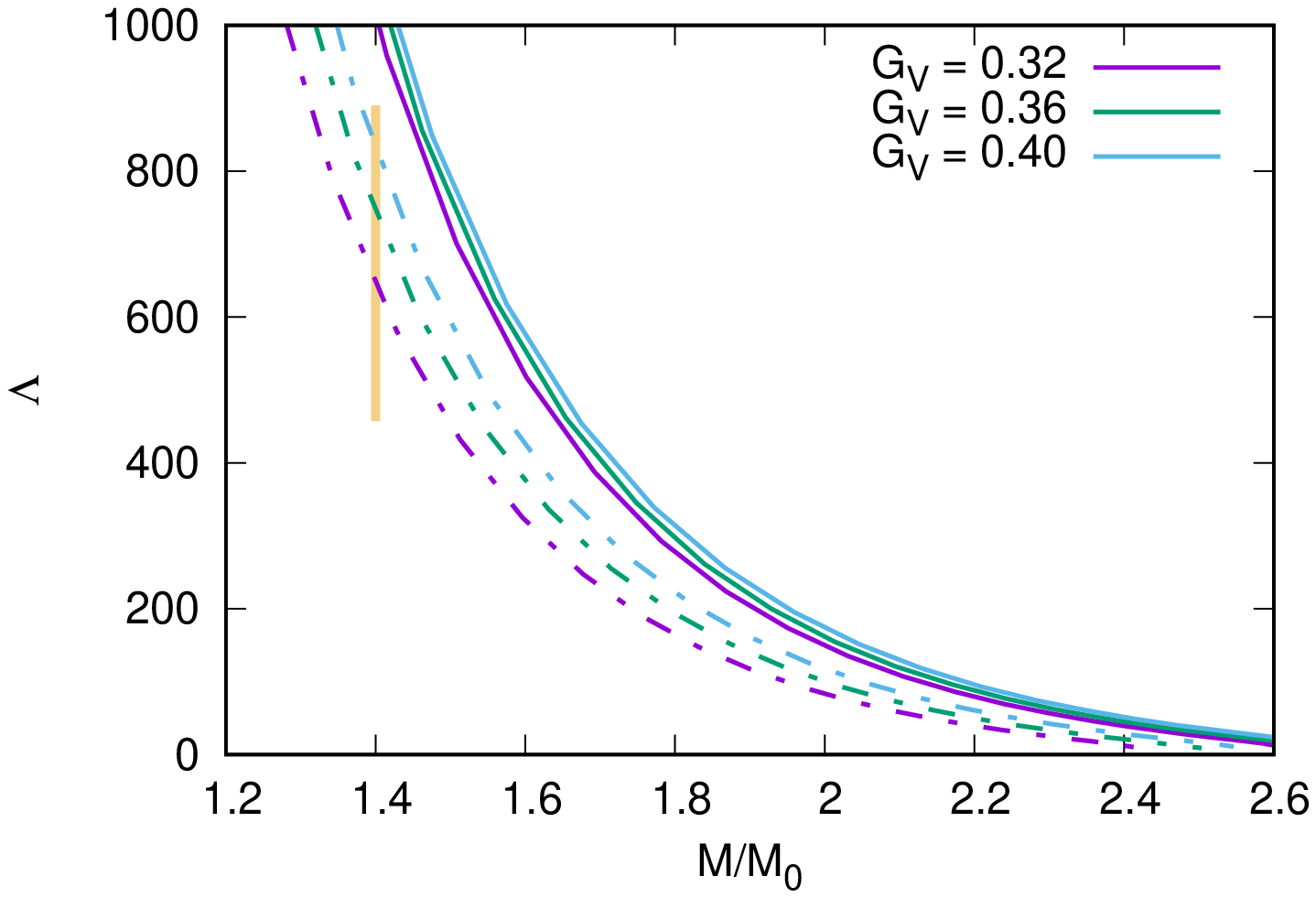} \\
\includegraphics[width=0.333\textwidth,,angle=270]{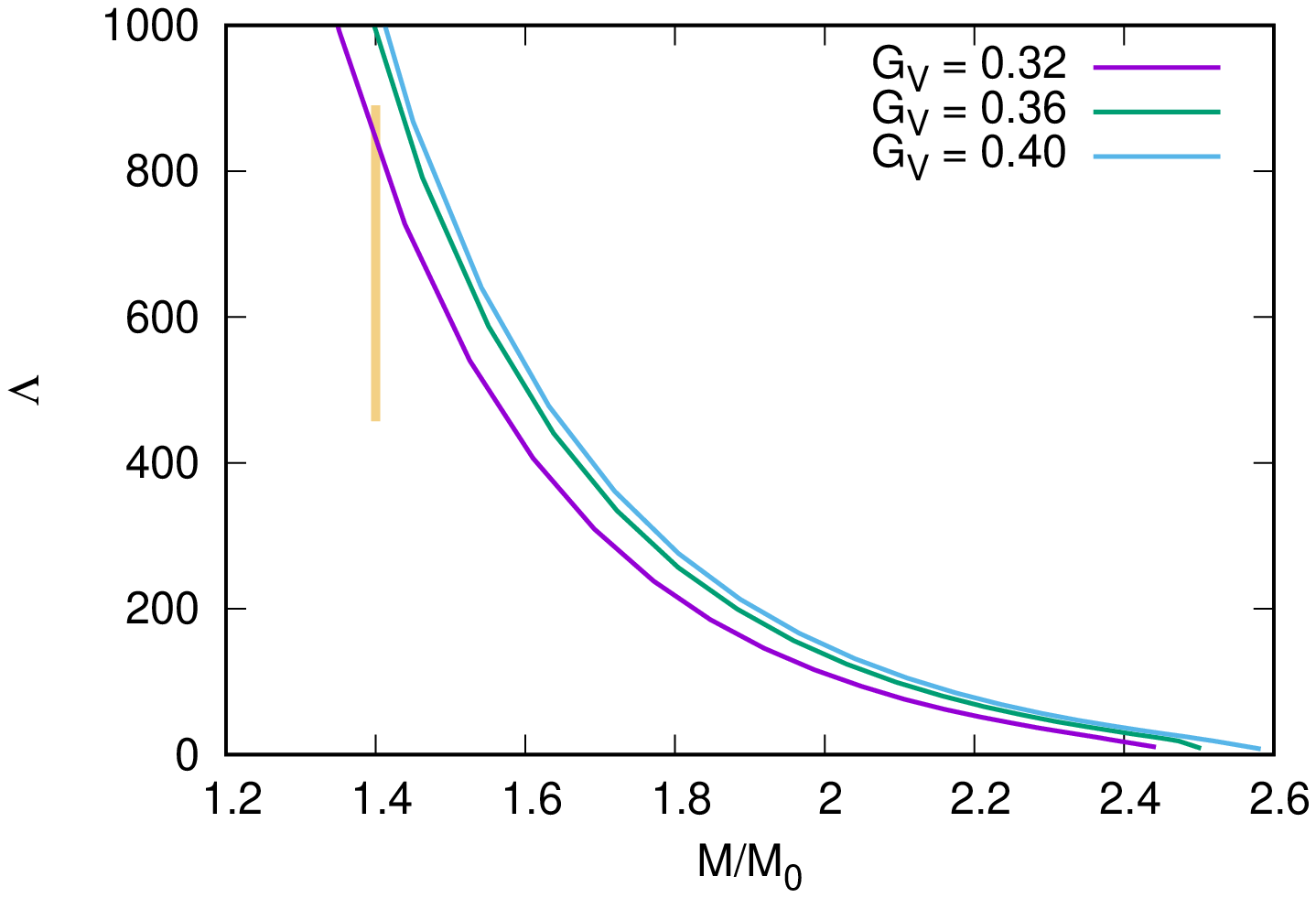} \\
\end{tabular}
\caption{(Color online) Dimensionless tidal parameter $\Lambda$ for strange stars with $X_V$ = 1.0 (top) and $X_V$ =0.4 (bottom).
Solid lines indicate minimum bag value for the stability window and  dotted lines indicate maximum bag value for the stability window.} \label{FL8}
\end{centering}
\end{figure}

As can be seen for $X_V$=0.4, only $G_V$ = 0.32 fm$^2$ fulfills the constraint for $\Lambda_{1.4}$ coming from
ref.~\cite{GW190814}. However, since its maximum mass is below 2.50 M$_\odot$, we can
rule out the  possibility of the mass-gap object of the  GW190814
event being a strange star with $X_V$ =0.4. In the case of $X_V$=1.0,
only the values of the maximum allowed bag 
fulfill the constraint. However, as pointed out earlier, these values
present  low radii, in disagreement 
with ref.~\cite{GW190814}. We cannot completely rule out the
possibility of the mass-gap object of the  GW190814  being a strange
star, since the radii values still fulfill NICER
constraints. Nevertheless, in order to keep internal 
coherence, in the light of the constraints presented in
ref.~\cite{GW190814}, we assert that the probability of the mass-gap
object being a strange star is significantly lower than that of being a hadronic star.

\begin{figure}[ht]
  \begin{centering}
\begin{tabular}{c}
\includegraphics[width=0.333\textwidth,angle=270]{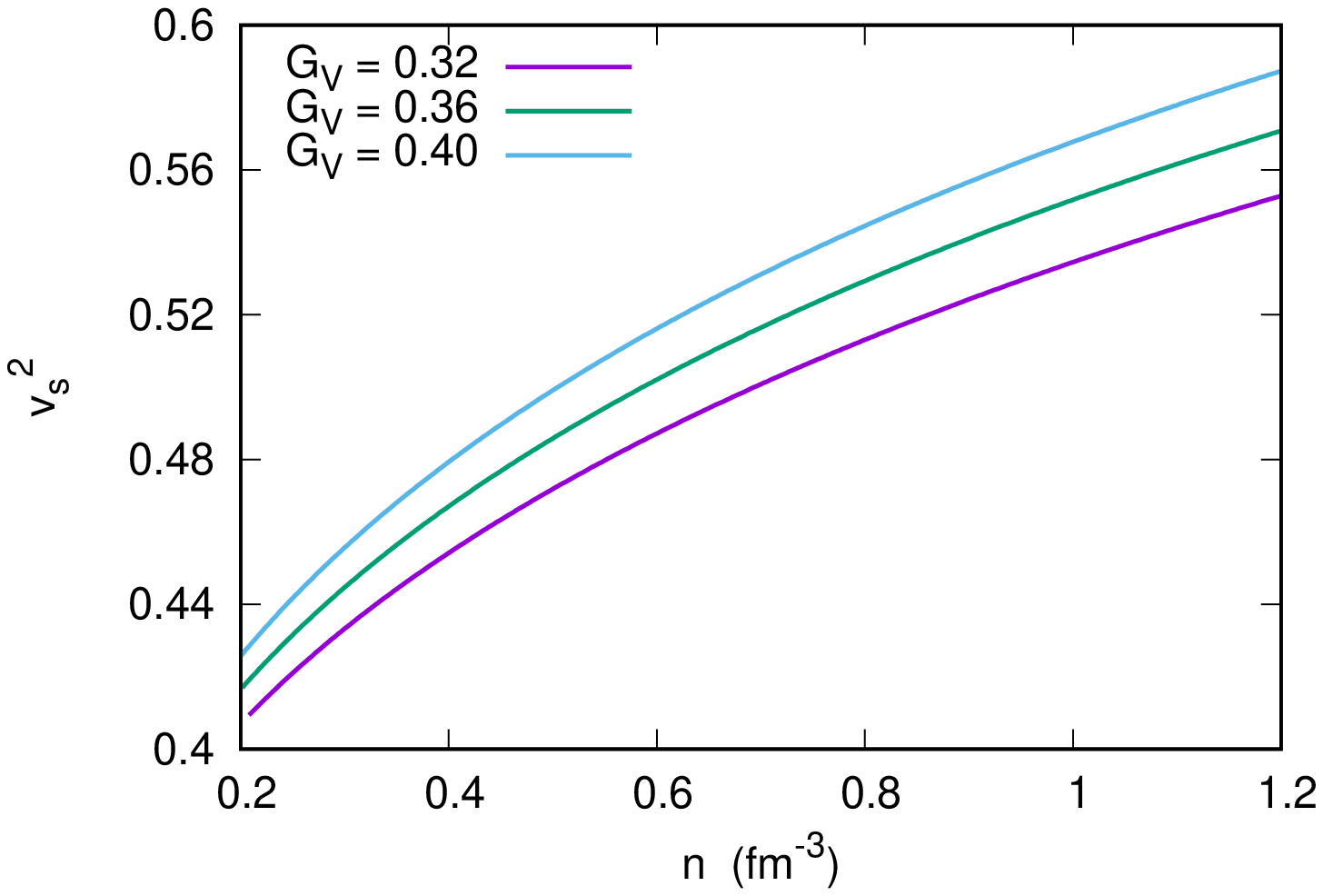} \\
\includegraphics[width=0.333\textwidth,,angle=270]{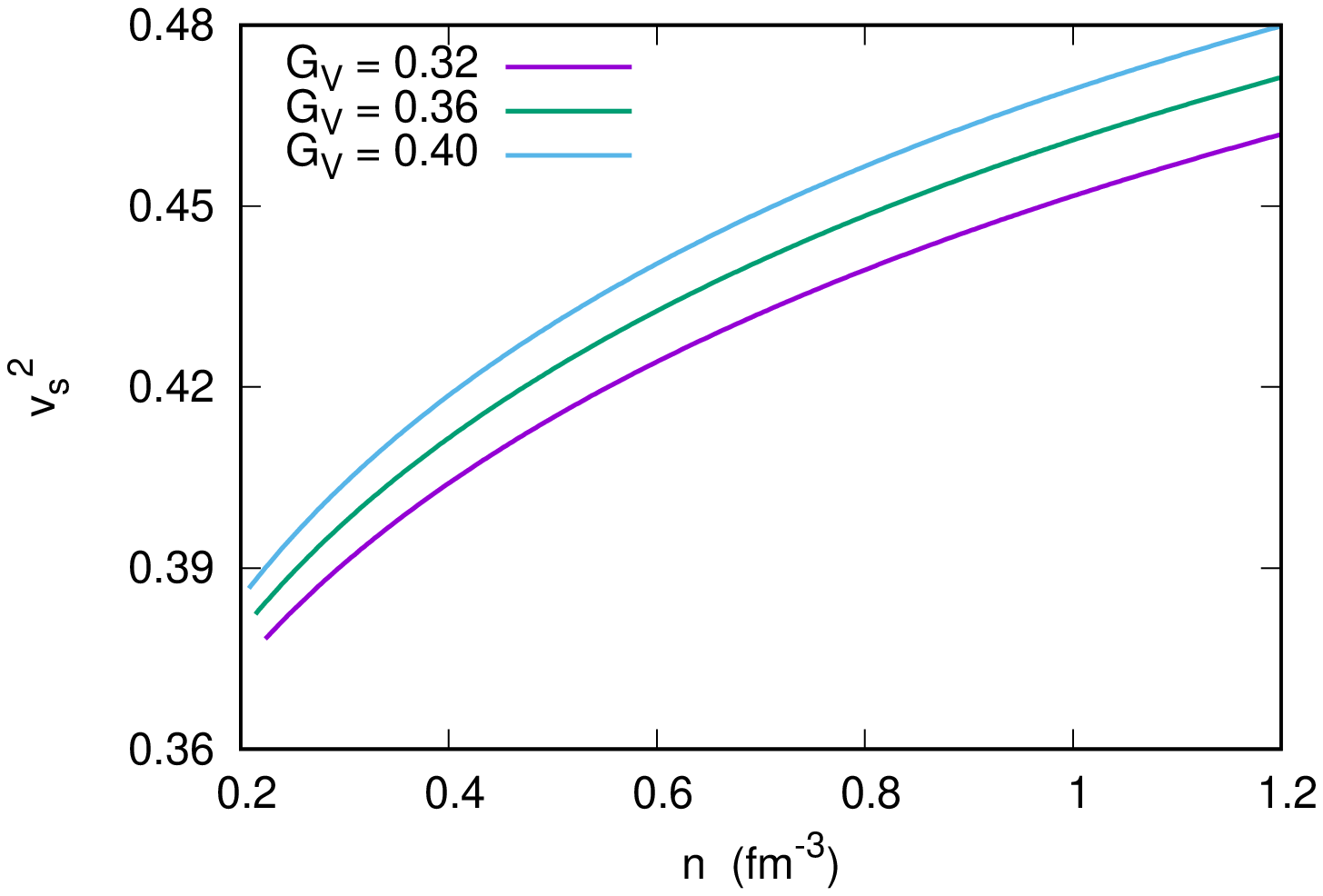} \\
\end{tabular}
\caption{(Color online) The square of the speed of sound with $X_V$ = 1.0 (top) and $X_V$ =0.4 (bottom).
} \label{FL9}
\end{centering}
\end{figure}
 
In fig.~\ref{FL9}, we display the speed of the sound. As can be seen,  the causality $v_s^2~<1$ is always satisfied.
However, the conformal limit $v_s^2~<1/3$ is violated, even at low
densities. Nevertheless, we are far below the pQCD limit of $n>40~n_0$. We finish this section displaying some macroscopic and microscopic properties of the strange stars in Tab.~\ref{TL5}.
We only show results for $X_V$=1.0 as discussed in the text because
for $X_V$ = 0.4 no parametrization is able 
to reach a $ M > 2.50 $ M$_\odot$ and $\Lambda_{1.4} < 889$ simultaneously.

\begin{widetext}
\begin{center}
\begin{table}[hb]
\begin{center}
\begin{tabular}{|c|c|c|c|c|c|c|c|c|c}
\hline 
 $G_V$ (fm$^2)$ & $B^{1/4}$ (MeV) & $M_{max}/M_0$ & R (km) & $n_c$ (fm$^{-3}$) & $\Delta\epsilon_S$ (MeV/fm$^{3}$)  &  $v_{sc}^2$ & $R_{1.4}$ (km) & $\Lambda_{1.4}$    \\
 \hline
 0.32 & 138 & 2.64  & 13.14 & 0.699 & 180 & 0.50  & 12.17 & 983  \\
 \hline
0.36 & 138  & 2.70 & 13.30 & 0.671 & 174  & 0.51  & 12.31 & 1023  \\
  \hline
0.40 & 137 & 2.81   & 13.78 & 0.632 & 166 & 0.52 & 12.58 & 1072                             \\
 \hline
 \hline
 0.32 & 145 & 2.44   & 11.96 & 0.794 & 216 & 0.51 & 11.40 & 634                             \\
 \hline
 0.36 & 144 & 2.51   & 12.26 & 0.759 & 208 & 0.52 & 11.59 & 729                             \\
 \hline
 0.40 &143 & 2.58   & 12.57 & 0.725 & 198 & 0.53 & 11.80 & 817                             \\
 \hline
Constraints & Stability window & $>~2.50$ & - &   -  & - & $ <~ 1.0$ & $12.2-13.7$  & $458-889$   \\ 
\hline
\end{tabular}
 
\caption{ Some strange star properties for $X_V$ = 1.0 and some astrophysical constraints.} 
\label{TL5}
\end{center}
\end{table}
\end{center}
\end{widetext}
 
As can be seen, for the minimum bag value, the 
tidal parameter $\Lambda_{1.4}$ is close or even higher than 1000.
We  highlight here, that for $B^{1/4}$ = 144 MeV and 
$G_V$ = 0.36, we can produce a massive star with $M > 2.50$ M$_\odot$
with $\Lambda_{1.4} < 800$, agreeing with ref.~\cite{tidal1},
while the radius agrees with Bayesian analysis~\cite{Michael,Yuxi} and
the result presented in ref.~\cite{Annala}. 
Unlike the hadronic neutron star, we see that the central speed of the
sound is only weakly linked to the stiffness of the EoS, being almost constant for all models.

\section{Hybrid stars} \label{sec4}

We now investigate the possibility of the GW190814 event being
a dynamically stable hybrid star. There are some questions we try to
answer next: It is possible that the mass-gap object in the  GW190814
event is a hybrid star? How do hyperons affect the presence of quarks
in the neutron star core? What are the values of the chemical
potential at the phase transition point? What are the maximum and
minimum values of $G_V$ that produce a dynamically stable hybrid star?
What is the influence of the bag value? How does the factor $X_V$
influence the macroscopic properties? What is the stellar minimum mass
that presents a quark core? What are the size and the mass of the quark core in a stable hybrid star?

The possibility that the mass-gap object  of the  GW190814 event being
a hybrid star has already been studied in ref.~\cite{Han}. 
In that paper, the authors use the generic constant-sound-speed (CSS)
parametrization, with no information about the quark interaction nor
the chemical composition of the matter and the EoS  reads~\cite{generic}:

\begin{equation}
 p(\epsilon) = a(\epsilon - \epsilon_0) , \label{EL27}
\end{equation}
where $a$  and $\epsilon_0$ are constants related to the speed of the
sound and the energy density  at zero pressure, respectively.
Here we use a more physical Lagrangian density based model described 
in Eq.~(\ref{EL22}).

Another difference between the present work and the study of
ref~\cite{Han} is the quark hadron phase transition criteria.
In ref.~\cite{Han} the transition pressure is treated as a free
parameter. We use the so called Maxwell construction, 
and the transition pressure is the one where the Gibbs free
energy per baryon $G/n_B$ of both phases intersect, being the
energetically preferred phase the one with lower
$G/n_B$~\cite{Chamel}. The Gibbs free energy per baryon coincides with 
the baryon chemical potential, therefore we next call the intersection point as critical pressure and critical 
chemical potential. The Maxwell criteria read:

\begin{equation}
 \mu_H = \mu_Q = \mu_C, \quad \mbox{at} \quad p_H = p_Q = p_C , \label{EL28}
\end{equation}
where the subscript $H$ indicates a Hadronic phase, $Q$ indicates quark phase and $C$ indicates the critical values.

We begin by reanalysing the stability window of Fig.~\ref{FL6}. To construct strange stars, we imposed that the 
energy per baryon of the deconfined phase was lower  than the nonstrange infinite baryonic matter.
But here we need the opposite. To produce a stable hybrid star, the
strange matter must be unstable, otherwise, as soon as the core of the
star converts into the quark phase, the entire star will convert into
a quark star in a finite amount of time~\cite{Olinto,Kauan}. 
Therefore we fix our bag values between 150 MeV $<~B^{1/4}~<$ 160 MeV to ensure an unstable strange matter.
We study the possibility of a hybrid star with nucleons and quarks,
and a hybrid star with nucleons, hyperons and quarks.
To obtain very massive hybrid stars with nucleons and hyperons, we use
next $\alpha$ = 0.25. Now, for a chosen  value of $X_V$,
we solve the TOV equations for different values of $G_V$ and of the
bag. The values of $G_V$ that produce a stable hybrid star with $M >
2.50$ M$_\odot$ form what we call the hybrid branch stability window.

\begin{figure}[ht]
  \begin{centering}
\begin{tabular}{c}
\includegraphics[width=0.333\textwidth,angle=270]{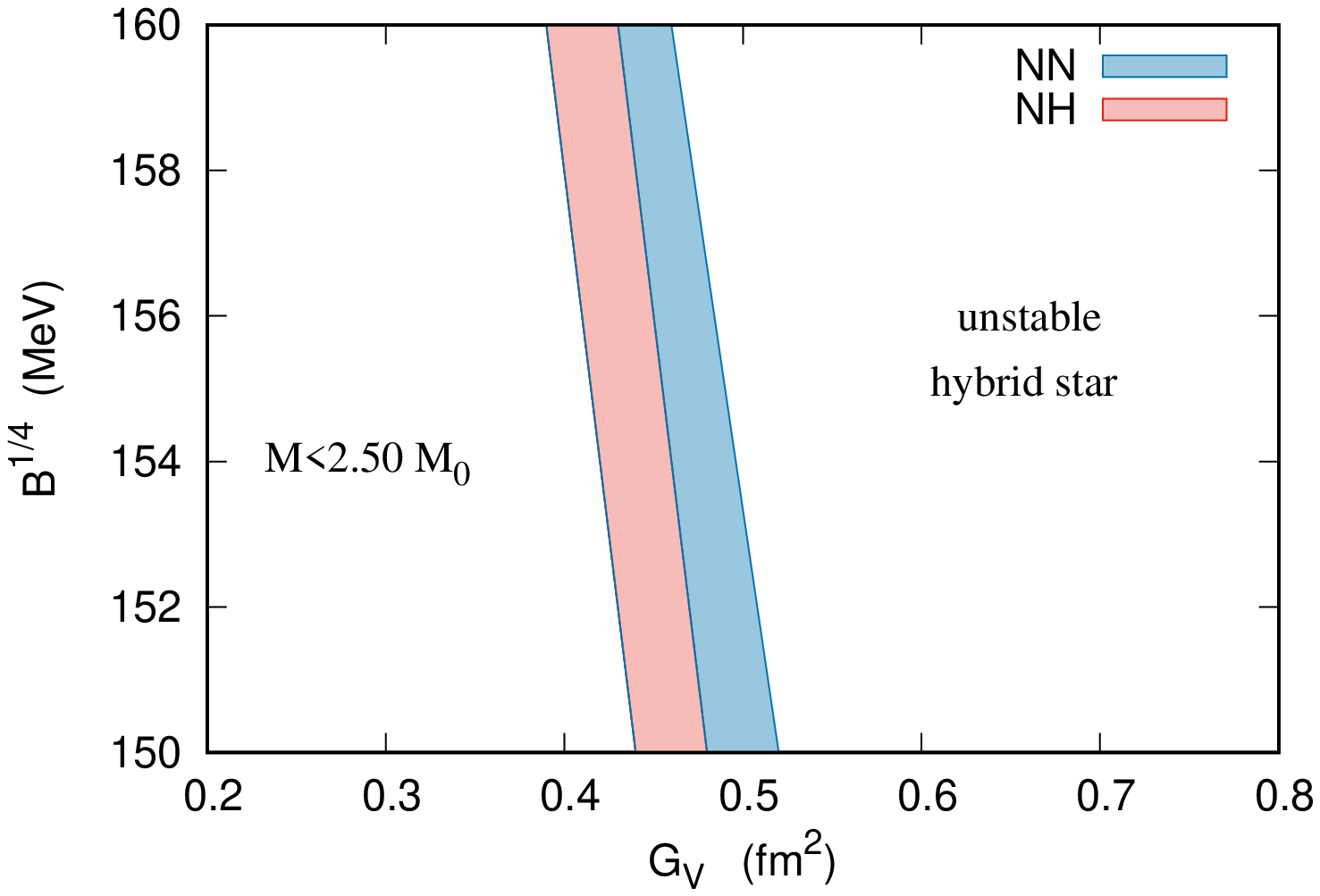} \\
\includegraphics[width=0.333\textwidth,,angle=270]{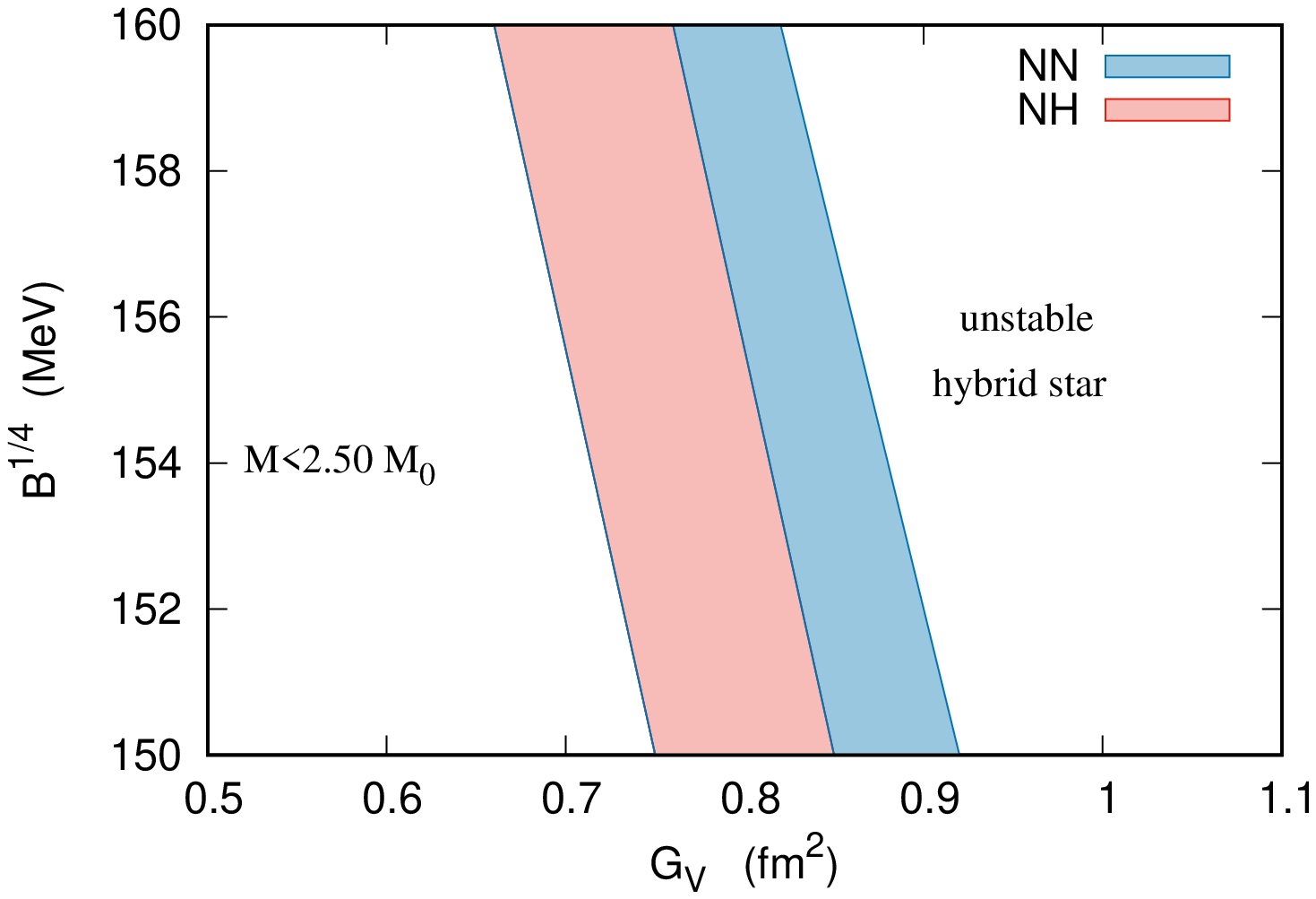} \\
\end{tabular}
\caption{(Color online) Hybrid branch stability window for $X_V$ = 1.0
  (top) and $X_V$ = 0.4 (bottom) with a pure nucleonic EoS (NN) and an
  EoS with nucleons and hyperons (NH).)
} \label{FL10}
\end{centering}
\end{figure}

\begin{figure*}[t]
\begin{tabular}{cc}
\centering % \begin{center}/\end{center} takes some additional vertical space
\includegraphics[scale=.51, angle=270]{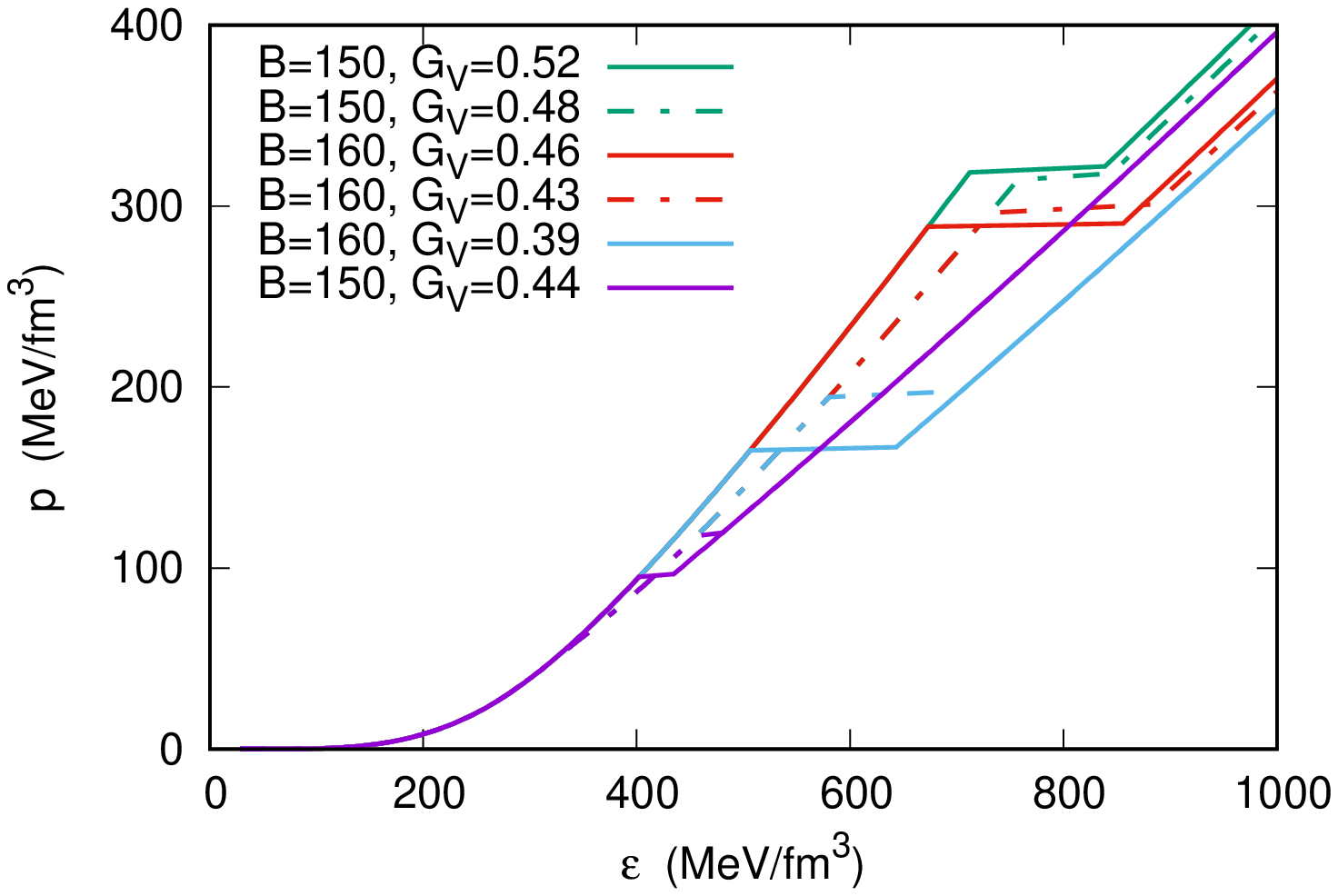} &
\includegraphics[scale=.51, angle=270]{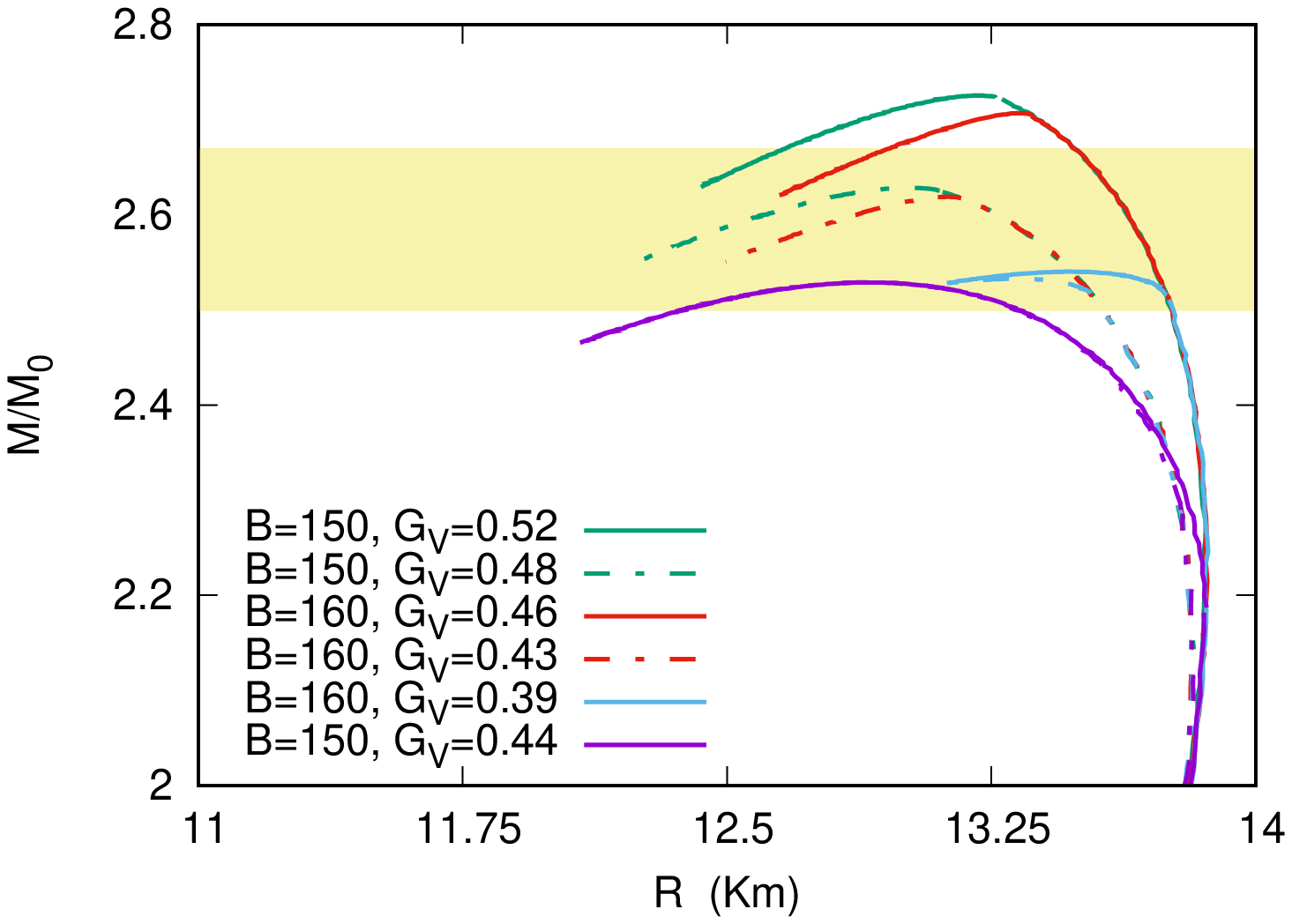}\\
\includegraphics[scale=.51, angle=270]{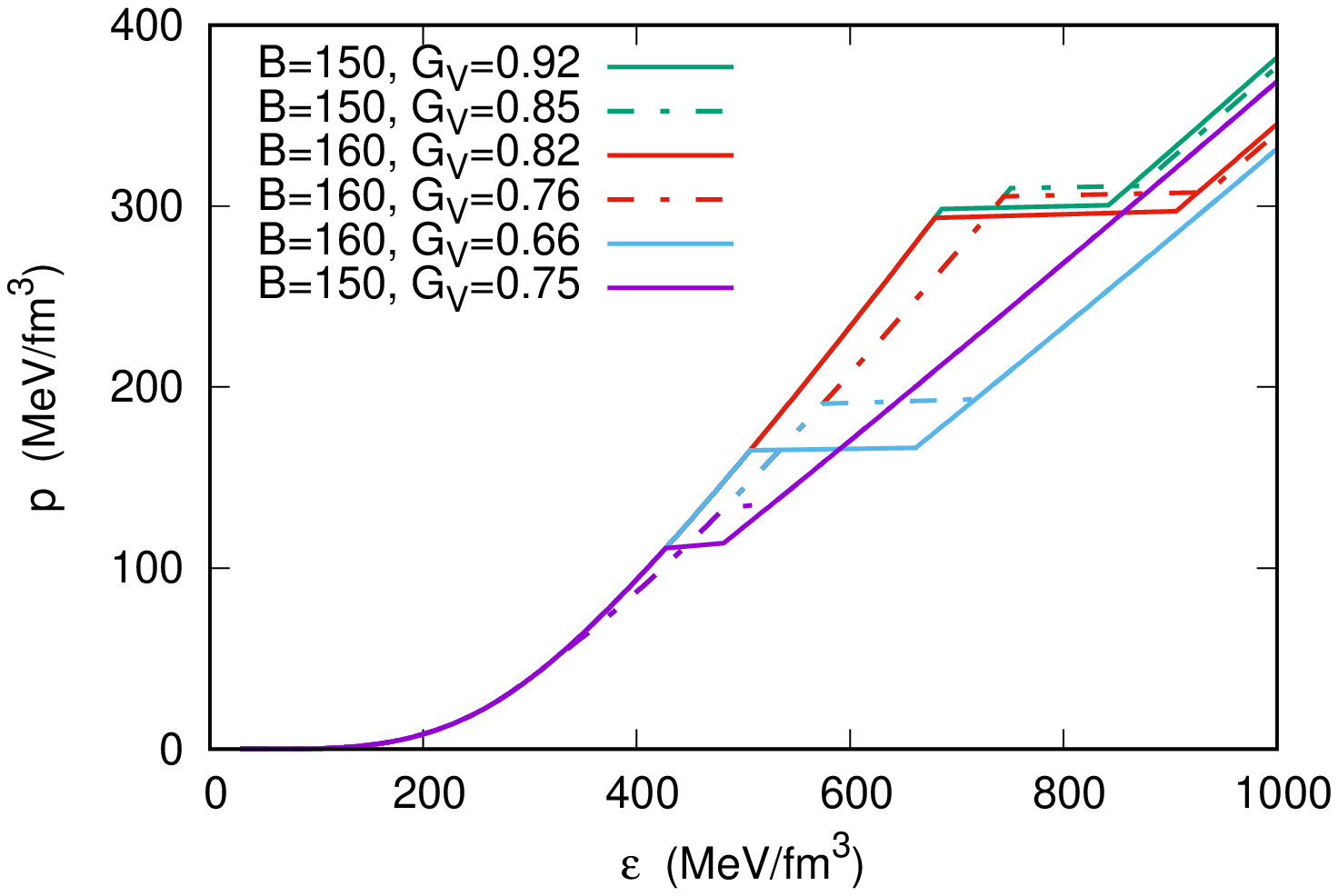} &
\includegraphics[scale=.51, angle=270]{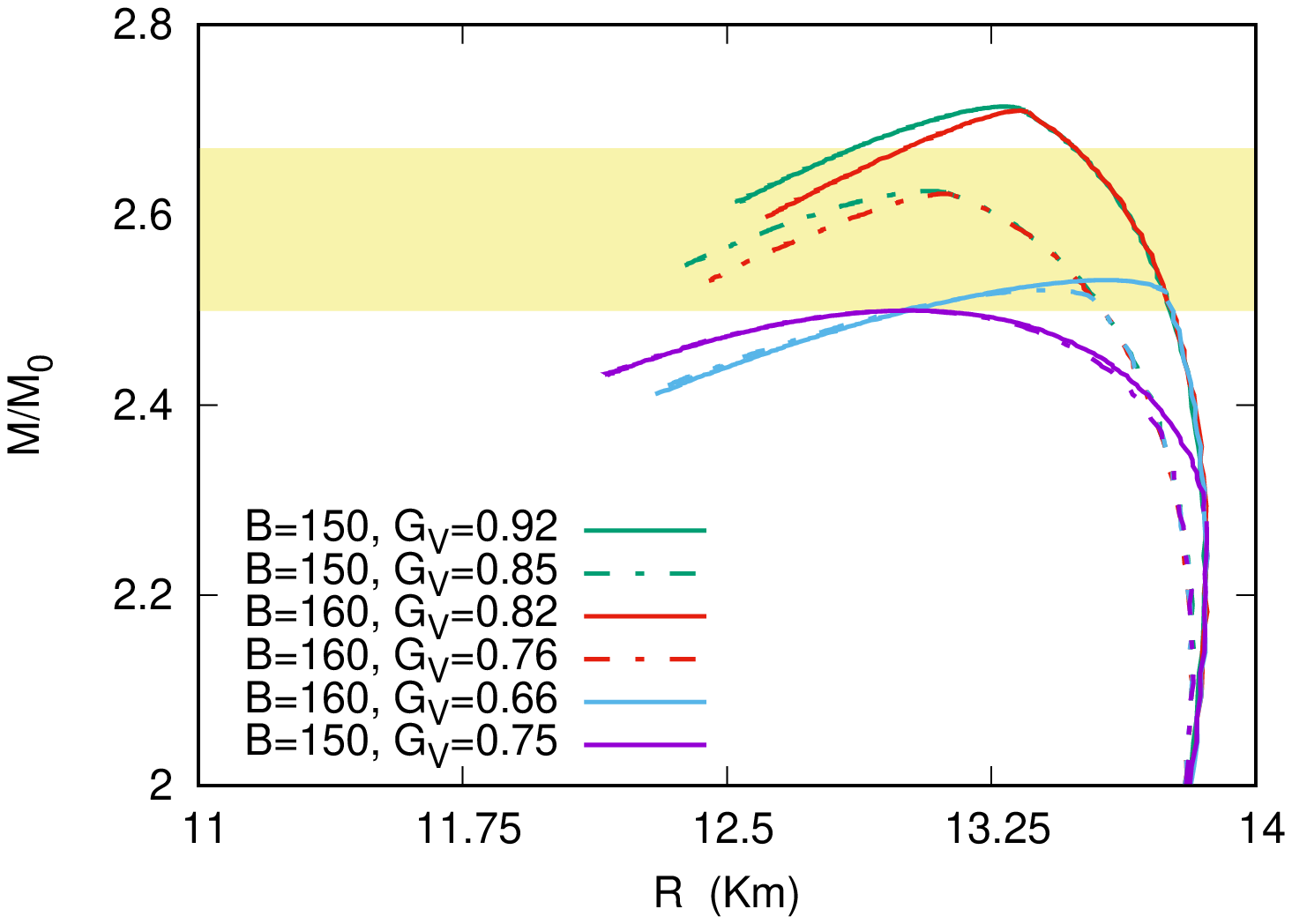}\\
\end{tabular}
% "\includegraphics" is very powerful; the graphicx package is already loaded
\caption{(Color online) EoS (left) and TOV solution (right) hybrid  stars with $X_V$ = 1.0 (top) and $X_V$ =0.4 (bottom) at the extreme values of $G_V$.
Solid lines indicate a hadronic phase with only nucleons (NN) and   dotted lines indicate a hadronic phase with nucleons and hyperons (NH)} \label{FL11}
\end{figure*}

We start by constructing the hybrid branch stability window for pure
nucleonic (NN) and hyperonic (NH) EoS 
for the hadronic phase with $X_V$ = 1.0 and $X_V$ = 0.4. The results are presented in Fig.~\ref{FL10}.
We can see that for $X_V$ = 1.0 and  150 MeV $<~B^{1/4}~<$ 160 MeV,
the values of $G_V$ that produce dynamical stable hybrid stars, with
$M > 2.50M_\odot$ lie between 0.39 fm$^2 < G_V < $ 0.48 fm$^2$ for a
hybrid star with nucleons, hyperons and quarks, and 0.39 fm$^2 < G_V <
$ 0.52 fm$^2$ for a hybrid star with only nucleons and quarks. With
$X_V$ = 0.4 the results are very different. The values of $G_V$ now
lie between 0.66 fm$^2 < G_V < $ 0.85 fm$^2$ for a hybrid star with
nucleons, hyperons and quarks and 0.66 fm$^2 < G_V < $ 0.92 fm$^2$ 
for a hybrid star with only nucleons and quarks. We also see that the hybrid
branch stability window for $X_V$ = 0.4 is  significantly broader 
when compared with $X_V$ = 1.0. The consequences 
of using a $G_V$ below or above the values of the  hybrid branch 
stability window are also different. If $G_V$ is to low, stable hybrid
stars still exist, however the maximum mass is lower than 2.50
M$_\odot$. If $G_V$ is too high, there is no 
dynamically stable hybrid star, yet the maximum mass is above 2.50 M$_\odot$, but it is purely hadronic.

We also estimate the mass and size of the quark cores present in the
most massive hybrid star of each model presented. To accomplish that, we follow ref.~\cite{lopesnpa,lopesmnras} and solve the TOV equations~\cite{TOV} for the quark EoS from the density corresponding to the critical chemical potential up to the density at the maximum mass. The EoS and the TOV solution with $X_V$ = 1.0 and $X_V$ = 0.4 for the extreme values of $G_V$ that satisfy the hybrid branch stability window are  shown in Fig.~\ref{FL11}, while the macroscopic and microscopic properties are displayed in Tab.~\ref{TL6}. 

Several different comparisons can be done from the results presented
 in Tab.~\ref{TL6}. For instance, for a fixed value of $B$, 
 $X_V$, and $G_V$ we can see the influence of hyperons in hybrid stars. In this case, the presence of hyperons does not affect the maximum mass of a hybrid star. Indeed, hyperons soften the
 EoS and make the hadron-quark phase transition more difficult. A
 hybrid star family with only nucleons present a lower critical
 chemical potential, therefore a lower $M_{min}$, which is the minimum
 star that presents a quark core, and at the same time present a 
larger quark core when compared with the case with hyperons.
 
 By increasing $G_V$ we see that an increase of the maximum mass. However, as it also increases the critical chemical potential, the $M_{min}$ becomes  very close to the maximum mass. For the maximum allowed value of $G_V$, the mass and radius of  the quark core are very small. Moreover,  we see that nucleonic hybrid stars allow a higher value of $G_V$, and therefore produce a higher maximum mass.
 
 For a correct choice of $G_V$, the maximum mass does not depend on the bag value. However, the mass and radius of the quark core do. Higher values of the bag produce lower sizes of the quark core, even for the lower value of $G_V$ in the  hybrid branch stability window.
  
The mass and size of the quark core strongly depend on the value of
the critical chemical potential. Higher values of $\mu_C$ produce low
values of the quark core.  Massive quark cores can be obtained with 
low values of the bag and low values of $G_V$. We can also see that the critical chemical potential can be as high as 1700 MeV. This value is significantly higher than those around 1200 MeV presented in other studies about hybrid stars~\cite{lopesmnras,Klahn2,Ayriyan}.

For a correct choice of $G_V$, the hybrid star maximum mass is
independent of $X_V$. However $X_V$ =0.4 seems to produce a slightly
lower value of the mass and radius of the quark core when compared
with the ones obtained with $X_V$ = 1.0.

The mass and radius of the quark core are strongly model
dependent. The mass can vary from only 0.002 
M$_\odot$ to values larger than 1 solar masses. The quark core can
vary from less than 1 km to almost 8 km. 
Also, we do not discuss the tidal deformability or the radius of the
canonical mass, once in our model all hybrid stars have a mass of at
least 2.18 M$_\odot$.

We finish our analysis discussing the speed of the sound at the
critical chemical potential in the light of the results presented in
ref.~\cite{Annala}. The authors claim that he speed of the sound of the quark matter is closely related to the mass and radius of the quark core in hybrid stars. Moreover, they assert that if the conformal bound ($v_s^2 <$ 1/3 ) is not strongly violated, massive neutron stars are predicted to have sizable quark-matter cores. As can be
seem, we do not find such correlation between the size of the quark
core and the speed of sound. We are even able to produce a quark core
with mass above 1 solar masses, with $v_s^2 =$ 0.50, far above the
conformal limit. Also, the most massive quark core is not those with
the lower speed of the sound.  

\begin{widetext}
\begin{center}
\begin{table}[ht]
\begin{center}
\begin{tabular}{|c|c|c|c|c|c|c|c|c|c|c|c|c}
\hline 
$X_V$  & $B^{1/4}$ (MeV)  & Type & $G_V$ (fm$^2)$  & $M_{max}/M_\odot$ & R (km)  & $n_c$ (fm$^{-3}$)  & $\mu_C$ (MeV) & $M_{min}/M_\odot$  & $M_Q$ ($M_\odot)$ & $R_Q$ (km) & $v_{SC}^2$ \\
 \hline
 1.0 & 150 & NH  & 0.44 & 2.53 & 12.90 & 0.719  & 1336 & 2.28 & 0.874 & 7.23 & 0.50 \\
 \hline
1.0 & 150  & NN & 0.44 & 2.53 & 12.90  & 0.720  & 1280 & 2.18 & 1.064 & 7.89 & 0.49 \\
    \hline
 1.0 & 150 & NH  & 0.48 & 2.63 & 13.02 & 0.710  & 1707 & 2.62 & 0.029 & 2.08 & 0.55  \\
 \hline
1.0 & 150  & NN & 0.52 & 2.73 & 13.21  & 0.689  & 1736 & 2.72 & 0.019 & 1.82 & 0.57 \\
  \hline
1.0 & 160 & NH   & 0.39 & 2.53 & 13.31 & 0.665 & 1499 & 2.51 & 0.125 & 3.55 & 0.51 \\
\hline
1.0 & 160 & NN   & 0.39 & 2.54 & 13.47 & 0.624 & 1445 & 2.50 & 0.148 & 3.87 & 0.51 \\
\hline
1.0 & 160 & NH   & 0.43 & 2.62 & 13.13 & 0.718 & 1680 & 2.62 & 0.009 & 1.35 & 0.54 \\
\hline
1.0 & 160 & NN   & 0.46 & 2.71 & 13.32 & 0.718 & 1683 & 2.70  & 0.043 & 2.36 & 0.55
\\
 \hline
 \hline
 0.40 & 150 & NH   & 0.75 & 2.50 & 13.00 & 0.712 & 1373 & 2.35 & 0.650 & 6.44 & 0.48                          \\
 \hline
 0.40 & 150 & NN   & 0.75 & 2.50 & 13.01 & 0.713 & 1321 & 2.27  & 0.817 & 7.08 & 0.47                          \\
 \hline
 0.40 &150 & NH   & 0.85 & 2.63 & 13.08 & 0.715 & 1695 & 2.62  & 0.015  & 1.66 & 0.51                        \\
 \hline
 0.40 & 150 & NN   & 0.92 & 2.71 & 13.29 & 0.689 & 1703 & 2.71 & 0.011 & 1.50 & 0.51
\\
\hline
 0.40 & 160 & NH   & 0.66 & 2.52 & 13.40 & 0.661 & 1490 & 2.51  & 0.087 & 3.14   & 0.48                        \\
 \hline
 0.40 & 160 & NN   & 0.66 & 2.53 & 13.57 & 0.632 & 1444 & 2.51  & 0.115 & 3.53  & 0.48                         \\
 \hline
 0.40 &160 & NH   & 0.76 & 2.62 & 13.11 & 0.740 & 1689 & 2.62 & 0.002 & 0.85 & 0.50                        \\
 \hline
 0.40 & 160 & NN   & 0.82 & 2.71 & 13.33 & 0.715 & 1695 & 2.71 & 0.004 & 1.11 & 0.51
\\
\hline
\end{tabular} 
\caption{ Maximum mass, radius, central density, critical chemical potential, $M_{min}$,  mass and radius of the quark core and the speed of sound at the critical chemical potential for the extreme values of $G_V$ that allows dynamical stable hybrid stars. } 
\label{TL6}
\end{center}
\end{table}
\end{center}
\end{widetext}

\section{Conclusion}

In this work we discuss the possibility of the mass-gap object in the
GW190814 event being a degenerate object instead of a %more probable 
black hole. The main remarks are resumed as follows:

\begin{itemize}
 
  \item We start by looking if there is a parametrization of the QHD
    that fulfills all symmetric nuclear matter constraints, as
    discussed in two review works (ref.~\cite{Dutra2014,Micaela2017})
    and, at the same time, produce stars that can reach at least 2.50
    M$_\odot$. We choose a modified version of the NL3* to accomplish this task.
  
  \item We then study in what conditions hyperons can be present in
    such massive neutron stars. We show that a hyperonic neutron star
    with mass above 2.50 M$_\odot$ is possible if we use symmetry
    group arguments to fix the hyperon-meson coupling constant and
    use $\alpha <$ 0.75.
  
  \item We show that the maximum mass, as well as the strangeness
    fraction  are strongly linked to the choice of $\alpha$. The lower the value of $\alpha$, the lower the strangeness fraction, and the higher  the maximum mass.
  
  \item The maximum mass is also linked with the speed of the sound at
    the core of the neutron star. The lower  the $\alpha$, the higher
    the speed of sound and, consequently the higher the maximum
    mass. The conformal limit ($v_s^2 <$ 1/3 )~\cite{Paulo}, is always violated.
  
  \item We discuss the minimum mass that enables DU process, in the
    light of the constraint M$_{DU} > 1.50M_\odot$~\cite{DU}. We see
    that only pure nucleonic neutron stars are able to undergo a DU
    process and only for stars with M$ > 2.56M_\odot$.
  
  \item We also analyse the constraints related to the canonical M =
    1.40 M$_\odot$ star. As hyperons are only present in stars with M$
    > 1.66 $ M$_\odot$, all hadronic models present the same $R_{1.4}$
    = 13.38 km and $\Lambda_{1.4}$ = 644.   These values agree with
    the main constraint from ref.~\cite{GW190814}, i.e., 
12.2 km $< R_{1.4} <$ 13.7 and 458 $< \Lambda_{1.4} <$ 889.  
  
  \item We then study the possibility of the mass-gap object being a
    self-bound quark star, satisfying the Bodmer-Witten
    conjecture~\cite{Bod,Witten}. We construct the stability window
    for two values of $X_V$ and show that for $X_V$ = 0.4, no strange
    star is able to simultaneously fulfill M$ > 2.50 $ M$_\odot$ and $\Lambda_{1.4} < 889$.
  
  \item For $X_V$ = 1.0 we are able to simultaneously fulfill M$ >
    2.50 $ M$_\odot$ and $\Lambda_{1.4} < 889$, but with $R_{1.4} <$
    12.2 km. This is in disagreement with the main constraint, but
    still in agreement with others results found in the literature~\cite{NICER1,Michael,Yuxi}.
  
  \item We finally analyse the possibility that the mass-gap object is
    a hybrid star.  We discuss if such hybrid star can be composed by
    nucleons, hyperons and quarks, or by nucleons only and quarks.
We show that for a correct choice of $G_V$, both possibilities can be
satisfied. Also, we are able to produce hybrid stars for all values of the bag lying between 150 MeV and 160 MeV, for both $X_V$ = 1.0 and $X_V$ = 0.4
  
  \item For a fixed $G_V$, the presence of hyperons reduces the mass
    and radius of the quark core, but has little effect on the maximum  mass of the hybrid star.
  
  \item The size and the mass of the quark core is strongly model
    depend, its mass vary from values lower than 0.01 $M_\odot$ to values larger than 1.0 $M_\odot$.
  
  \item We did not find a correlation between the speed of the sound
    of the quark matter and the size of the quark core, as suggested
    in ref.~\cite{Annala}. Nevertheless, we are able to produce a very
  massive quark core M$ > 1$ M$_\odot$, even though the conformal limit is violated.

\end{itemize}

{ Acknowledgments}: This  work is a part of the project INCT-FNA
Proc. No. 464898/2014-5. 
D.P.M. is partially supported by Conselho Nacional de Desenvolvimento Cient\'ifico e  Tecnol\'ogico (CNPq/Brazil) under grant 301155.2017-8.

\end{document}